\documentclass{aastex63}

\received{XXX yyy, ZZZZ}
\revised{xxx1 yyy1, ZZZ1}
\accepted{October 19, 2021}

\submitjournal{AJ}

\shorttitle{Ages and Masses of M33 Star Clusters}
\shortauthors{Moeller and Calzetti}

\graphicspath{{./}{figures/}}
\DeclareUnicodeCharacter{2212}{-}

\begin{document}

\title{Ages and Masses of Star Clusters in M33: a Multi--wavelength Study}

\correspondingauthor{Caitlin Moeller}
\email{cmoeller@utexas.edu, calzetti@astro.umass.edu}

\author[0000-0002-0786-7307]{Caitlin Moeller}
\affiliation{University of Massachusetts Amherst \\
710 N Pleasant St \\
Amherst, MA 01003, USA}
\affiliation{University of Texas at Austin \\
23 San Jacinto Blvd \\
Austin, TX 78712, USA}

\author[0000-0002-5189-8004]{Daniela Calzetti}
\affiliation{University of Massachusetts Amherst \\
710 N Pleasant St \\
Amherst, MA 01003, USA}

\begin{abstract}
We combine archival images for the nearby galaxy M33 (Triangulum Galaxy) from the ultraviolet (UV) to the infrared (IR) to derive ages, masses, and extinctions for the young star cluster population, and compare our physical parameters with published ones. Our goal is to test the robustness of clusters’ ages and masses, and possibly improve on existing ones both by expanding the wavelength range of the spectral energy distribution (SED) fits and by using more recent population synthesis models. The rationale for this experiment is to verify the sensitivity of the clusters’ physical parameters to observational setups and model choices that span those commonly found in the literature. We derive the physical parameters of 137 clusters, using SEDs  measured in eight UV--to--I bands, including H$\alpha$, from \textit{GALEX} and ground--based images. We also add the 24~$\mu$m image from the \textit{Spitzer Space Telescope} to help break some age degeneracies. We find that our derived cluster ages show significant differences with earlier determinations, while the masses remain relatively insensitive to the fitting approach adopted. We also highlight an already known difficulty in recovering old, low--extinction clusters, as SED fitting codes tend to prefer younger, higher extinction solutions when the extinction is a free parameter. We publish updated ages, masses, and extinctions, with uncertainties, for all sample star clusters, together with their photometry; given the proximity of M33, this represents an important population to secure for the study of star formation and cluster evolution in spirals.
\end{abstract}

\keywords{stars: formation --- 
galaxies: star clusters: general ---  --- }

\section{Introduction} \label{sec:intro}

Within galaxies, the structures of star formation are a continuous, scale--free hierarchy from parsecs to
kiloparsecs \citep{lada2003, Elmegreen2003, Bressert+2010}, which are expected to arise from
the self--similar distribution of a turbulence--dominated ISM \citep{Elmegreen+1997}, mediated by
magnetic fields and outflow feedback \citep[e.g.,][]{Krumholz+2019}. Star clusters form in the dense regions of the hierarchy \citep[e.g.,][]{Elmegreen2010}, and, thus, provide a sensitive probe of the star formation process, while the rest of the hierarchy is quickly dispersed into the stellar field of the galaxy, over timescales of a few tens of Myr \citep{bastian2008, Longmore2014, grasha2015, grasha2017a}. Young star clusters contain the majority of the massive stars, which drive the feedback into the surrounding medium and regulate star formation \citep{Krumholz+2019, Adamo+2020}, but most are not bound because they are not dense enough \citep[e.g.][]{BrownGnedin2021}. These clusters are dispersed and randomized over short timescales ($\lesssim$10~Myr) by a vast array of processes, both internal-- such as gas expulsion, stellar evolution, and two–body relaxation,-- and external,-- such as tidal shear, random motions, interactions with molecular clouds, and secular evolution of the host galaxy \citep[e.g.,][]{GielesBastian2008}. 

Deriving accurate ages and masses for star clusters is a key step for linking the small scales of the star formation, i.e., the scales of individual stars, with the large scales of whole galaxies. The physical and chemical properties of star clusters constrain scenarios for their formation and evolution, trace the star formation history of galaxies, constrain models of cloud collapse, provide tests for the origin and persistence of spiral arms, and guide cosmological simulations \citep{Shabani+2018, Li+2018, Adamo+2020,  BallesterosParedes+2020}. The distribution of cluster masses is a sensitive tracer of formation mechanisms \citep[e.g.][]{whitmore1999, larsen2002, gieles2006a, gieles2006b, whitmore2014, johnson2017, Adamo+2020}, while the age distribution reveal the cluster disruption mechanisms \citep[e.g.,][]{Gieles2009, Bastian2012, SilvaVilla+2014, chandar2016, Adamo+2017, messa2018b}. 

M33 (Triangulum Galaxy) is an excellent laboratory for testing techniques for the derivation of physical parameters of star clusters. This nearby spiral galaxy \citep[850 kpc;][]{Ferrarese+2000} is a member of the Local Group and has a large population of clusters. M33 is close enough that its star clusters can be resolved into stars with the \textit{Hubble Space Telescope}, but distant enough that lower angular resolution facilities can survey it fully. Its modest inclination ($\sim$52$^o$, from  NED\footnote{NED=NASA Extragalactic Database,  http://ned.ipac.caltech.edu/}) minimizes the chance of line--of--sight confusion. At the M33 distance, an angular aperture of  1$^{\prime\prime}$ radius subtends a 4.1$\times$5.2~pc$^2$  spatial region, which is comparable to the size of star clusters \citep{ryon2017, BrownGnedin2021}. \citet{Sarajedini2007} compiled a list of compact  sources found in M33, for a total of 451 objects, 255 of which were confirmed to be star clusters. For these clusters, ages and masses were determined in a series of papers by Ma et al. \citep{Ma+2001,  Ma+2002b, Ma+2002c, Ma+2002a, Ma+2004a, Ma+2004b}, using spectral energy distribution (SED) fitting of medium-band optical filter photometry. 

\begin{figure}
 \centering
 \includegraphics[width=12cm]{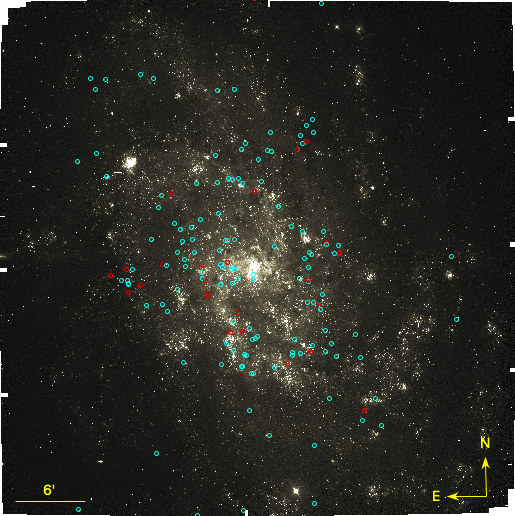}
    \caption{Central pointing view of M33 galaxy \citep[U band, from][]{Massey+2006} showing 154 of the 163 star clusters from \citet{Sarajedini2007} analyzed in this work. The remaining nine clusters are located in the Northern and Southern pointings, and are not shown here. A scale of 6$^{\prime}$ correspond to $\sim$1.5~kpc  at the distance of this galaxy. Cyan circles mark clusters for which SED fits were performed, while red circles mark  clusters excluded from the  fits, and for which only photometry is provided. See text for more details.}
        \label{regions}
\end{figure}

Starting from this compilation, we combine   broad--band optical imaging with the two UV  \textit{GALEX} images, add the hydrogen recombination line H$\alpha$, and include the 24~$\mu$m band map from the \textit{Spitzer Space Telescope}, all from archival holdings. Our goal is to leverage the extended wavelength coverage of our photometric measurements to produce accurate ages and masses for the cluster population in this iconic galaxy. The UV is a sensitive discriminator of young versus old populations and, combined with the optical bands, adds a leverage arm for extinction measurements \citep{Calzetti+2015}, while H$\alpha$ tracks the presence of ionizing massive stars. Finally, the 24~$\mu$m band traces the emission from the dust heated by the UV photons of young stars \citep[$\lesssim$100~Myr,][]{KennicuttEvans2012}. Thus the 24~$\mu$m emission enables discriminating intermediate age populations ($\sim$10-100~Myr) when multiple solutions are possible but the H$\alpha$ emission is no longer present and the UV is mostly yielding degenerate results \citep{Leitherer+1999}. The metal abundance of M33 is sufficiently high that we expect significant dust and 24~$\mu$m emission in correspondence of its young star clusters. \citet{U+2009}, \citet{Bresolin+2011} and \citet{Toribio+2016} report values between 1/2 solar and $\sim$20\% above solar\footnote{We adopt a solar oxygen abundance of 8.69, \citep{Asplund+2009}} for both the young stars and the gas within the inner $\sim$3.5~kpc of this galaxy (0.4 R$_{25}$), where most of our sources are located (Figure~\ref{regions}). We compare our results with those of Ma et al. to investigate similarities and differences. 

The outline of the paper is as follows. Section 2 describes the cluster compilation of \citet{Sarajedini2007}. Section 3 discusses the archival imaging data, our processing and photometry. Section 4 describes the models employed for SED fitting and our fitting approach. Section 5 presents the results of the SED fits and our analysis, including comparisons with previous determinations of  ages and masses. Section 6 summarizes our findings and conclusions. We present our updated ages and masses and extinctions, as well as photometry, in tabular form for all star  clusters in our sample. 
  
\section{Cluster Sample} \label{sec:identification}

The clusters in our sample come from a compilation of \citet[][(henceforth SM2007)]{Sarajedini2007} of 451 compact sources in M33, among which the authors identified 255 star clusters using high angular resolution imaging from either the \textit{Hubble Space Telescope} or ground--based facilities. Of these 255 confirmed star clusters, SM2007 lists both ages and masses for a total of 163, as published in a series of papers by Ma et al. \citep{Ma+2001, Ma+2002b, Ma+2002c, Ma+2002a,   Ma+2004a, Ma+2004b}. Ma et al. derived ages and masses by fitting the population synthesis models of \citet[][with the 1996 improvements]{BruzualCharlot1993} to the SEDs of the sources. The SEDs consisted of photometry in 8 intermediate--band filters in the wavelength range 3800--10,000~\r{A}\  \citep{Ma+2001}. This wavelength range does not enable the authors to constrain uniquely the internal dust reddening of the star clusters, which they either adopt from previous analyses---as in \citet{Ma+2001}, where the authors use the extinction values of \citet{Chandar+1999}---or fix E(B--V) to a constant value \citep[E(B--V)=0.1,][]{Ma+2002b}. The values of the ages and masses are given without uncertainties. In their SED fits, Ma et al. allowed  the metallicity to remain a free parameter, ranging between 3\% solar and 1.4 times solar, with the reference solar metallicity being Z$_{\odot}$=0.02 \citep{Iglesias+1992}.

Although optical photometry for the sources is  published in SM2007, we do not use these authors' measurements as we aim to derive uniform--aperture photometry for the star clusters, for the purpose of using  internally--consistent measures for our SED fitting experiment.

\section{Imaging Data and Photometry} \label{sec:data}

\subsection{The Data}
Ultraviolet, optical and IR images from a combination of space and ground-based imaging data of the M33 galaxy were retrieved from NED, for a total of 10 bands: Far--UV (FUV), Near--UV (NUV), U, B, V, R, I, H$\alpha$ (6563~\r{A}), 3.6$\mu$m and 24$\mu$m. The FUV and NUV images trace the youngest stars, with ages $\lesssim$100~Myr, while the ionized gas emission at H$\alpha$ traces even younger stars, $\lesssim$7--10~Myr, although this age limit increases by about a factor of 3 in the presence of binary stars \citep{Xiao+2018}. The IR image at 24~$\mu$m traces the emission of the dust heated by the young stars \citep{DraineLi2007, Calzetti+2007}.  

The \textit{GALEX} satellite \citep{Martin+2005} observed M33 in both the FUV ($\sim$1530~\r{A}) and NUV ($\sim$2310~\r{A}) channels with a single pointing of its $\sim$1.6$^o\times$1.6$^o$ field-of-view (FoV) with angular resolution $\sim$4$^{\prime\prime}$--5$^{\prime\prime}$\citep[e.g.,][]{Dale+2009}. The optical band images were obtained by \citet[][for U, B, V, R, and I]{Massey+2006} and \citet[][for H$\alpha$]{Massey+2007} at the Mayall 4--m telescope at KPNO, with the Mosaic Camera, with a 38$^{\prime}\times$38$^{\prime}$ FoV and angular resolution between 0$^{\prime\prime}$.8 and 1$^{\prime\prime}$. To observe the entirety of M33, images were taken in three different pointings: Northern, Central, and Southern. Mosaics at the wavelengths 3.6~$\mu$m and 24~$\mu$m were obtained with the \textit{Spitzer Space Telescope} IRAC and MIPS cameras, respectively, with angular resolution of 1$^{\prime\prime}$.7 at 3.6~$\mu$m and 6$^{\prime\prime}$.5 at 24~$\mu$m \citep{Fazio+2004, Rieke+2004}.

All images were aligned and resampled to the pixel scale of the optical ones (0$^{\prime\prime}$.27 per pixel), to preserve the highest possible angular resolution. This facilitates the identification of the location of the SM2007 sources onto the images, and the assessment of whether neighboring sources are present which may affect the ages and mass determinations. The latter step is made necessary by the significantly lower resolution of the \textit{GALEX} and \textit{Spitzer}/MIPS images relative to the ground--based ones.

\subsection{Photomery}
Multi-wavelength photometry was performed on the images using the positions published in the SM2007 catalog. An aperture of 3$^{\prime\prime}$ in radius was adopted as a compromise between enclosing as much as possible of the UV and IR point spread functions (PSFs) while at the same time minimizing the overlap between photometric apertures of neighboring star clusters. An outer annulus of 1$^{\prime\prime}$.1 width around each aperture was used to estimate the background and remove it from the aperture measurement. A consequence of the small aperture size is to exclude the wings of the UV and IR PSFs, which leads to an  underestimate of the flux of the sources in these bands. We correct for this effect by applying aperture corrections derived by studying the growth curves of isolated sources in the images, and listed in Table~\ref{corr_table}. The optical broad--band photometry does not require aperture corrections.

We do not apply aperture corrections to the narrow--band photometry, although it is evident from visual inspection that the ionized gas emission is sometimes more extended than the size of our photometric aperture. This aperture subtends  a physical radius of 12~pc at the distance of M33, which corresponds to the Str\"omgren radius of a 10$^4$~M$_{\odot}$, 5~Myr old star cluster, for an assumed electron density n$_e\sim$20~cm$^{-3}$. In general, our clusters are either less massive or older, but some overfill the aperture with their ionized gas emission. In order to offset this limitation, after running the SED fits, we visually inspect the continuum--subtracted H$\alpha$ image in correspondence of each cluster to ensure that the age results are not driven by an underestimated H$\alpha$ flux.

\begin{table}[!htb]
\centering
\begin{tabular}{@{}cc@{}}
\toprule
Filter    & Correction Factor \\ \hline
FUV       & 2.30          \\
NUV       & 2.36          \\
3.6$\mu$m & 1.22          \\
24$\mu$m  & 5.00          \\ \hline
\end{tabular}
    \caption{Aperture correction factors applied to the photometry at the indicated wavelengths.}
    \label{corr_table}
\end{table}

The units of the UV and optical images are count--rates and counts, respectively. We convert those units to physical units of flux density (erg~s$^{-1}$~cm$^{-2}$~\r{A}$^{-1}$) using: the stellar photometry of \citet{Massey+2006} with the zeropoints published in \citet[p.~100]{Zombeck+1990} in the case of the \textit{UBVRI} images, the stellar photometry of \citet{Massey+2007} with the zeropoints published in the KPNO Mosaic instrument manual\footnote{https://www.noao.edu/kpno/mosaic/manual/mosa3\textunderscore 1.html} in the case of the narrowband H$\alpha$ image, and the conversion factors published in the \textit{GALEX} instrument manual\footnote{https://asd.gsfc.nasa.gov/archive/galex/Documents/instrument\textunderscore summary.html} in the case of the UV images. The list of flux conversion factors is in Table~\ref{f_conv_table}. Measured photometry for all confirmed 163 star clusters is listed in Table~\ref{photometry}. Fluxes are given {\em prior} to correction for foreground  Milky Way extinction, which amounts to E(B--V)=0.036 \citep{Schlafly+2011}. 

To directly observe the presence of line emission in our images, we subtract the stellar continuum as traced by the R--band image from the narrow--band filter targeting the H$\alpha$ line. We rescale the R image prior to subtraction, with the scaling factor derived by computing a ratio of the counts in R to the counts in the narrow--band for emission--line--free stellar sources. 
The narrow--band filter has a FWHM=80.62~\r{A}\ \citep{Massey+2007} and includes the two [NII](6548,6584~\r{A}) metal lines. We use the ratio [NII]/H$\alpha$=0.27 published in \citet{Kennicutt+2008} to remove the [NII] contamination from the narrow--band filter and obtain an H$\alpha$ image. 
The stellar continuum subtracted H$\alpha$ images are used solely for visual comparison to confirm the ages of the younger star clusters ($<$10~Myr) through the presence or absence of excess H$\alpha$ emission. The original, un-subtracted H$\alpha$ images are used for photometry and SED fitting.

The archival \textit{Spitzer} images are already calibrated in surface brightness units of MJy/sr. We convert those to fluxes (erg~s$^{-1}$~cm$^{-2}$) within our apertures using standard conversion factors, with the appropriate pixel scale of 0$^{\prime\prime}$.27 per pixel. The 3.6~$\mu$m photometry is further multiplied by the factor 0.91 to include photometric corrections\footnote{https://irsa.ipac.caltech.edu/data/SPITZER/docs/irac/iracinstrumenthandbook/46/}. The IRAC 3.6~$\mu$m image is only used to remove the underlying stellar emission from the 24~$\mu$m image, via the formula: F$_{24,dust}$[Jy]=F$_{24}$[Jy] - 0.035 F$_{3.6}$[Jy] \citep{Helou+2004, Calzetti+2007}. The photometry in these two bands is reported in Table~\ref{photometry} in units of flux. We use the 24~$\mu$m image mainly to break age degeneracies in the SED fits when the cluster is sufficiently old that the ionized gas emission is no longer present, but still young enough to have UV emission heating the dust. 

Visual inspection of all the measured star cluster reveals that 26 out of 163 cannot be retained  within our sample for SED fitting for a variety  of reasons, as detailed in Table ~\ref{exclusions}. The most common reason is the presence of multiple sources with different colors (likely different ages) within the 3$^{\prime\prime}$ radius aperture used for the photometry. We thus limit the SED fits to the remaining 137 sources for which  we are reasonably confident that one single source is contained within the photometric aperture, or, in case of multiple sources, these show similar colors (likely similar/equal ages).

\begin{table}[!htb]
\centering
\begin{tabular}{@{}cc@{}}
\toprule
Filter    & Flux Conversion Factor     \\ \hline
FUV       & 1.4E-15                    \\
NUV       & 2.06E-16                   \\ 
U         & 2.87E-21                   \\
B         & 7.75E-21                   \\
V         & 4.92E-21                   \\
R         & 2.39E-21                   \\
I         & 6.67E-22                   \\
H$\alpha$ & 7.13E-21                   \\ \hline
\end{tabular}
    \caption{Flux conversion factor used in each filter at UV and optical wavelengths, from either count--rate or counts to erg~s$^{-1}$~cm$^{-2}$~\r{A}$^{-1}$. See text for details.}
    \label{f_conv_table}
\end{table}

\subsection{Measurement Uncertainties}
The images at U, B, V, H$\alpha$, R, and I are sufficiently deep that flux calibration uncertainties dominate the error budget. These are taken from \citet{Massey+2006} for the broad--band filters and from \citet{Massey+2007} for the  H$\alpha$ filter. The uncertainties are listed in Table~\ref{logf_err_table} as uncertainties on the logarithm of the flux. For the optical filters, the value quoted is the largest uncertainty of the sample in each filter.

Uncertainties in the FUV, NUV, and in the (stellar continuum--subtracted) 24~$\mu$m photometry are dominated by uncertainties in the aperture placement, because of the large PSF size. We calculate these uncertainties by `wiggling' the position of the aperture by one pixel (0$^{\prime\prime}$.27) in  each direction and remeasuring the flux within the aperture. The scatter that results is used in the final error budget, listed in Table~\ref{logf_err_table}.

\begin{table}[!htb]
\begin{center}
\begin{tabular}{@{}cc@{}}
\toprule
Filter  & $\sigma$(logFlux) \\ \hline
U  & 0.011 \\
B  & 0.012 \\
V  & 0.018\\
R  & 0.019 \\
I  & 0.008 \\
H$\alpha$ & 0.04 \\
FUV & 0.22 \\
NUV & 0.15 \\
3.6$\mu$m & ... \\
24$\mu$m & 0.13 \\ \hline
\end{tabular}
\end{center}
    \caption{For the optical filters, we quote the calibration error from \citet{Massey+2006} and \citet{Massey+2007};  the value shown is the largest uncertainty for clusters in the luminosity range of our sample; most star clusters have smaller uncertainty in the optical. UV and 24$\mu$m flux errors are calculated through variations of the location of the photometric aperture's center, which dominate the uncertainty budget.}
    \label{logf_err_table}
\end{table}

\section{Models and Fitting Approach} \label{models}

\subsection{Models}
The stellar population synthesis models used by \citet{Ma+2001} and subsequent papers, i.e., the  models of \citet{BruzualCharlot1993} with the 1996 updates, are no longer available and did not include nebular emission. More modern versions exist, such as those of \citet{BruzualCharlot2003} and more recent ones, which also include several updates to the stellar populations characteristics \citep[e.g.][]{Gutkin2016}. \citet{Wofford+2016} compared the most recent versions of these models with those generated by Starburst99 \citep{Leitherer+1999, Vazquez+2005} with the addition of nebular lines from YGGDRASIL \citep{Zackrisson+2011}, and the BPASS ones \citep{Stanway+2020}, finding that all models yield similar result, with low dispersion in the median values of age, mass, and extinction. 

In this analysis, we adopt the Starburst99+YGGDRASIL simple stellar population (SSP) models, with the assumption that star clusters can be reasonably approximated by an instantaneous burst (single) population \citep{Wofford+2016}. The models are deterministic, meaning that the stellar Initial Mass Function (IMF) is assumed to be fully sampled from the lowest to the highest stellar masses. This is not the case for star clusters with masses$<$3,000--5,000~M$_{\odot}$ \citep[e.g.,][]{Cervino+2002}, implying that the age and mass determinations for clusters below this mass limit will carry an uncertainty larger than the formal one we quote. While the correct approach for such low mass clusters is to use a Bayesian fitting method that returns full posterior probability distributions for a stochastically  sampled IMF \citep[e.g., SLUG][]{Krumholz+2019b}, this approach is beyond the scope of our current work.

Starburst99 \citep{Leitherer+1999} spectral synthesis models were generated with a \citet{Kroupa2001} IMF in the range 0.1–120~M$_{\odot}$ and metallicity Z = 0.02. This  metallicity value is higher than the mean value of  the M33 region where most of our clusters are located, $\sim$0.5--1 solar, but we choose it for the following three reasons: (1) the model `ingredients' are inconsistently sampled for metallicity below solar \citep{Vazquez+2005}; (2) the next model metallicity available is about 1/2 solar, at the lower end of the range for our clusters; and (3) ages and masses derived via multi--band SED fits are relatively insensitive to variations of a $\sim$2.5 factor in metallicity of the models, as shown in Figures~\ref{B-VAge} and \ref{NUV-I}  \citep[see, also,][and SM2007]{deGrijs2005}. We do expect star clusters older than a few 100 Myr to have lower metallicity than younger ones, as also found by \citet{Ma+2001}. However, our study is mostly interested in star clusters younger than $\approx$1--3$\times$10$^8$~yrs, which is the age range where we expect the  addition of UV, H$\alpha$, and dust emission (24~$\mu$m) tracers to  have the most impact on the clusters' physical parameters determination. 

The models utilize the Padova tracks with AGB treatment, to better represent intermediate and old stellar populations  \citep{Girardi+2000, Vazquez+2005} and are generated for the age range 1 Myr–13 Gyr; the age steps are 1 Myr in the 1–15 Myr range, 10 Myr in the 20–100 Myr range, 100 Myr in the 200–1000 Myr range, and 1 Gyr in the 2--13 Gyr range. The addition of nebular emission lines via YGGDRASIL \citep{Zackrisson+2011} includes a covering factor of 50\% for the gas emission, to account for leakage of ionizing photons out of HII regions \citep[e.g.,][]{Calzetti+2021}. The models assume non--rotating, single stars. There is, however, increasing evidence that stellar population models implementing rotating, binary stars are better fits for data of star clusters, especially at young ages and low metallicities \citep{Stanway+2020}. Variations  at the young ages can be as large as a factor 2--3 \citep{Wofford+2016}. However, given that M33 star clusters span a range of ages much larger than this scatter and that the galaxy is fairly metal rich, we will not include effects of rotating and binary stars, although this should be noted as a point for future investigations. 
We note that the Starburst99+YGGDRASIL models do not include Very Massive Stars ($>$150~M$_{\odot}$) either, which are known to be present in young star clusters \citep[e.g.][]{Crowther+2010, Smith+2016} and are included in more recent populations synthesis models \citep[e.g. BPASS, ][]{Eldridge+2017, Stanway+2018}. Our goal in choosing the Starburst99+YGGDRASIL models is to remain close to the astrophysical assumptions used in the stellar population synthesis models used by \citet{Ma+2001} and subsequent papers.
 
The effects of dust on the stellar population SEDs are applied via a starburst attenuation curve \citep{Calzetti+2000}, with the equation:
\begin{equation}
    F(\lambda)_{mod,ext} = F(\lambda)_{model}10^{[-0.4E(B-V)k(\lambda)],}
    \label{fore_dust}
\end{equation}
where $k(\lambda)$ is the starburst curve, and E(B--V) is the color excess applied to the SEDs. We generate models with the color excess range E(B-−V) = 0–-1, with step 0.02. As we use the directly measured cluster photometry of Table~\ref{photometry}, the small foreground Milky Way extinction with E(B--V) = 0.036 is corrected with the Milky Way curve of \citet{Fitzpatrick1999}.

The dust-attenuated SEDs are then convolved with the transmission curve of each of the UV and optical filters to produce synthetic luminosities in 8 bands, that are normalized to the default mass of Starburst99, $10^{6}$ M$_{\odot}$. The synthetic photometry so obtained is then used to fit the observed SEDs of the star  clusters. 

\subsection{Fitting Approach}
\citet{Ma+2001} and subsequent works by these authors estimated the ages and masses of the star clusters in M33 by using a least square fit between observed and synthetic photometry \citep{Kong+2000}. We adopt the same approach, described by:
\begin{equation}
    \chi^{2}(n,t,E(B-V),M) = \sum_{i=1}^{8} [F(\lambda_i, n)_{obs} - F(\lambda_{i},t,E(B-V),M)_{mod,ext}]^{2}/\sigma(\lambda_i, n)^2
    \label{Eq16}
\end{equation}

\noindent where, $F(\lambda_{i}, n)_{obs}$ is the observed flux density of the nth star cluster in the i--th band, $F(\lambda_{i},t,E(B-V),M)_{mod,ext}$ is the synthetic flux density of the models described in the previous section in the i--th band, and $\sigma(\lambda_i, n)$ is the uncertainty in the observed flux density; the model populations are a function of the age \textit{t}, the color excess \textit{E(B-V)}, and the cluster mass \textit{M}. The index \textit{i} runs from 1 to 8, which is the number of bandpasses in our SEDs. The goal is to minimize the reduced $\chi^2_{red}$=$\chi^2/(N_{freed}-1)$, where N$_{freed}$=5 is the number of degrees of freedom for 8 independent datapoints and 3 parameters. Uncertainties in the fits are obtained by investigating the distribution of $\chi^2_{red}$ values around the minimum, as described in \citet{Calzetti+2021}; we also study the $\chi^2_{red}$ distributions to ensure that our best fits are not the result of local minima.

\citet{Ma+2001} explored solutions for three different metallicities -- Z = 0.0004, 0.004, and 0.02, finding that older star clusters are better fit by models with lower metallicity. However, as already discussed above, these authors could not constrain the amount of intrinsic extinction in the clusters, due to the limited wavelength range of their SEDs, and either adopted a constant value of E(B--V)=0.1 or values derived by previous authors. Conversely, we employ a single metallicity value, but extend the wavelength range of the clusters' SEDs to better separate age from extinction. Our goal is to understand how changing the wavelength range of the SEDs and the fitting approach affects the values of ages $<$1--3$\times$10$^8$~yr, where the age--extinction degeneracy is generally pronounced. The inclusion of UV colors  provides, theoretically, a better handle of the effects of extinction, and, when accompanied by the U$-$B color, provides a sensitive discriminant for separating age from extinction. With the addition of the H$\alpha$ emission we can better identify clusters younger than $\sim$7--8~Myr, a regime where the above colors are often of inconclusive diagnostics \citep[see discussions in][]{Boquien+2009, Calzetti2013, Calzetti+2015}. Furthermore, We use the 24$\mu$m emission as a discriminant for multiple--age solutions, i.e., for the presence of multiple minima in the distribution of the $\chi^2_{red}$ values. For clusters with two viable ages with  significant difference (e.g. 8 Myr and 80 Myr), a visual check of the 24$\mu$m emission at the location of the cluster is used as the deciding factor. The 24~$\mu$m dust emission has been shown to follow closely the 8~$\mu$m emission in HII regions \citep{Relano2009}, and the latter has been shown to decrease dramatically with increasing age of the star cluster out to at least 300--400~Myr \citep{Lin+2020}. \citet{Sharma+2011} showed that most of  the 24~$\mu$m bright sources in M33 are generally younger than 10~Myr.

\section{Analysis and Results} \label{analysis}

\subsection{Ages, Masses, and Extinctions}

\begin{figure}[ht]
 \centering
 \includegraphics[width=10cm]{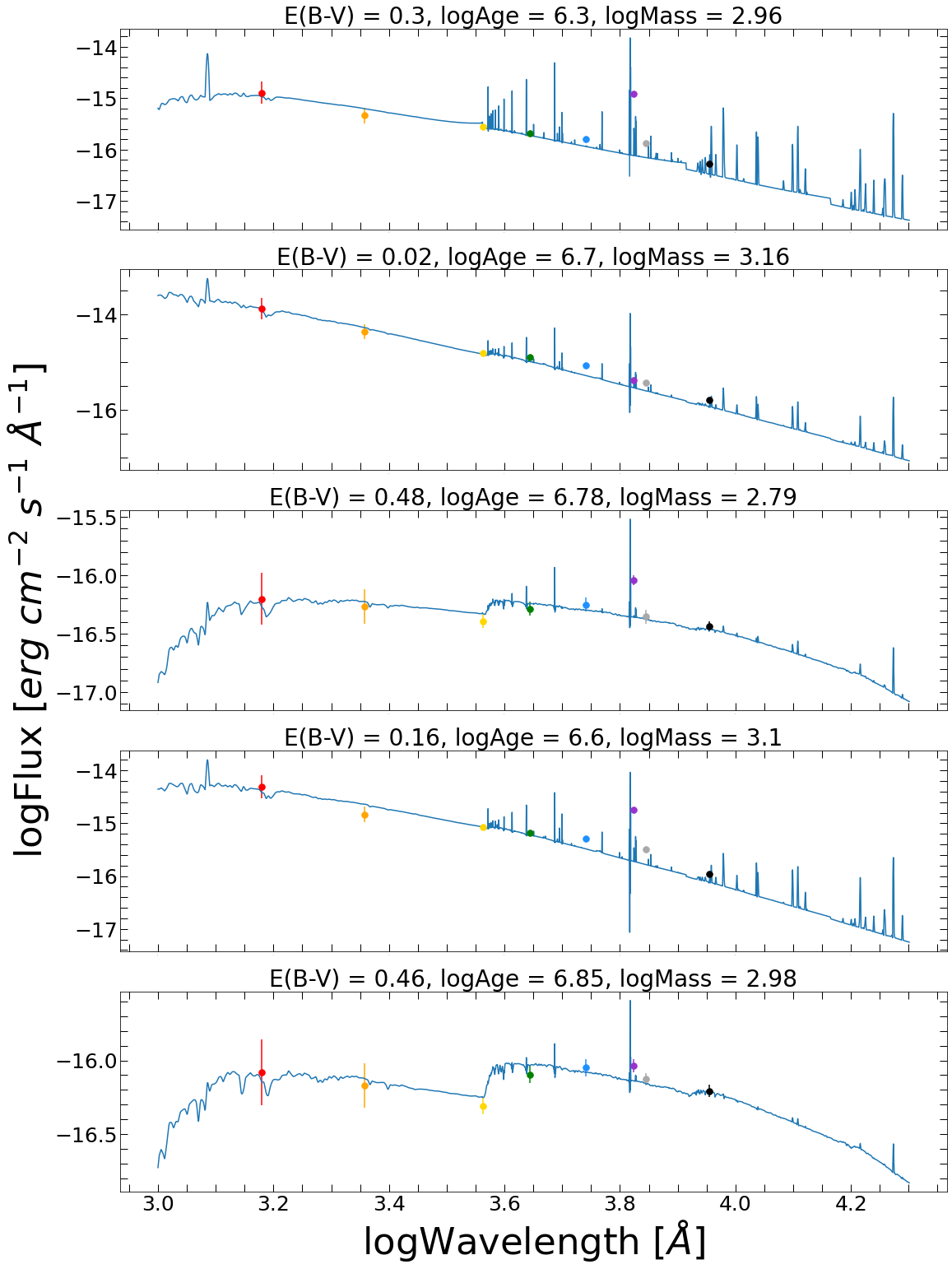}
    \caption{Five example SEDs of star clusters with $<$ 10 Myr. Beginning at the top plot, the clusters shown are 42, 70, 77, 79, and 111. The best fit extinction, age, mass of each cluster are  listed at the top of each panel, with  the age and masses in logarithmic scale. Units are:  mag for E(B--V), yr for ages and M$_{\odot}$ for masses.}
        \label{LessGallery}
\end{figure}

\begin{figure}[ht]
 \centering
 \includegraphics[width=10cm]{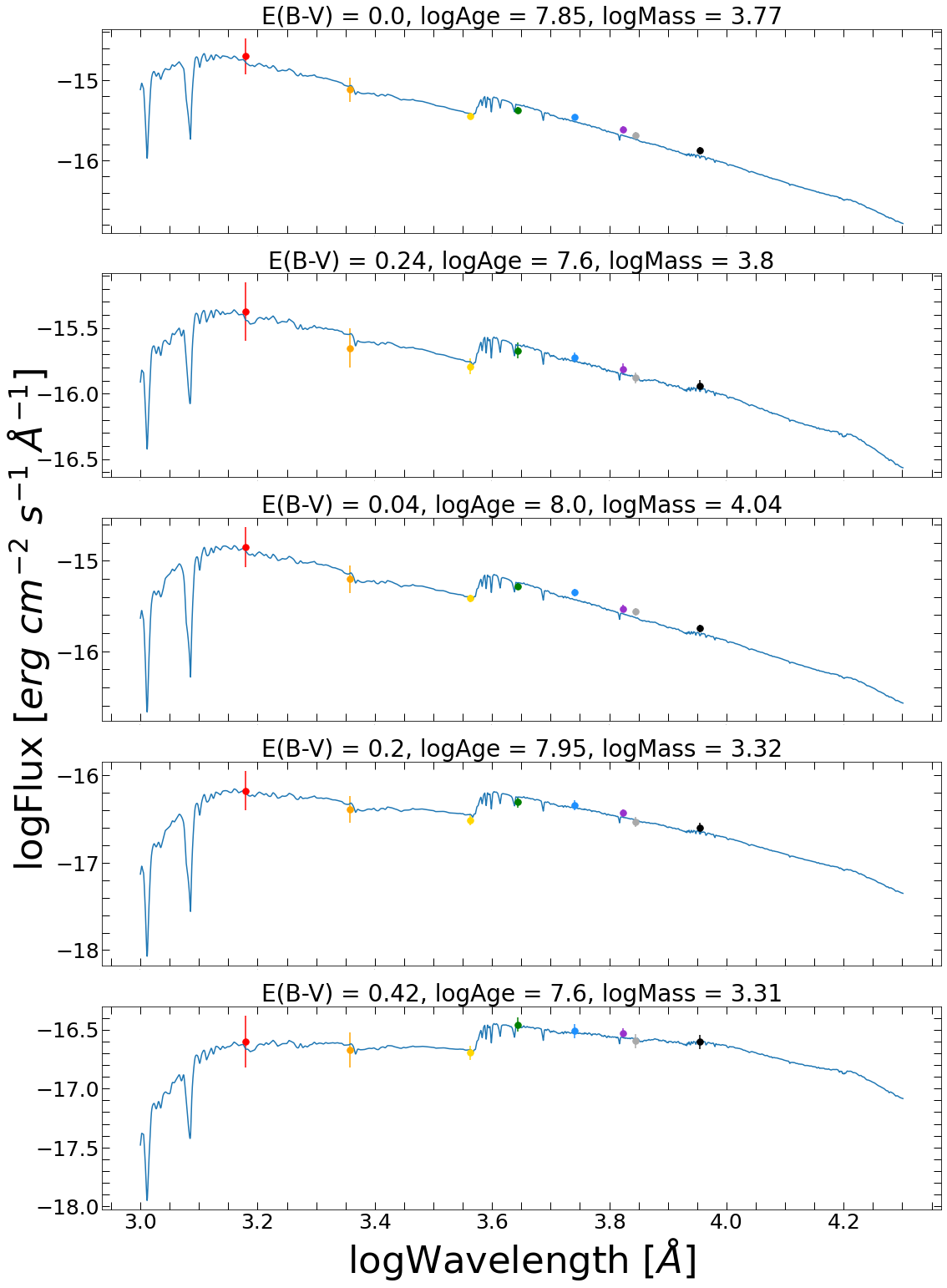}
    \caption{Five example SEDs of star clusters in the age range 10--100 Myr. Beginning at the top plot, the clusters shown are 22, 33, 52, 61, and 134. See caption of Figure~\ref{LessGallery} for more details. }
        \label{MidGallery}
\end{figure}

\begin{figure}[ht]
 \centering
 \includegraphics[width=10cm]{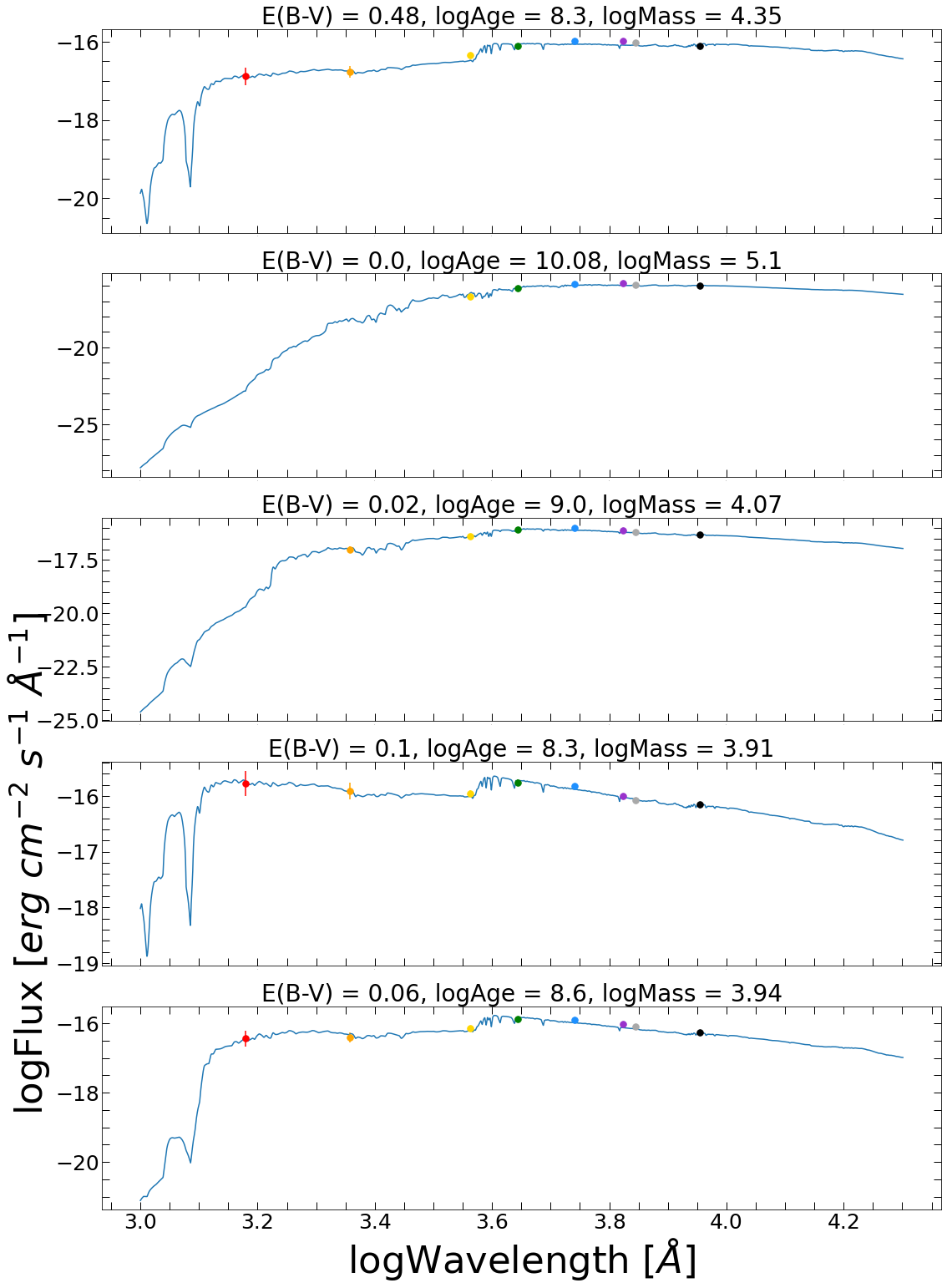}
    \caption{Five example SEDs of star clusters with $>$ 100 Myr age. Beginning at the top plot, the clusters shown are 24, 48, 80, 105, and 137. See caption of Figure~\ref{LessGallery} for more details.} 
        \label{MoreGallery}
\end{figure}

\begin{figure}[ht]
 \centering
 \includegraphics[width=15cm]{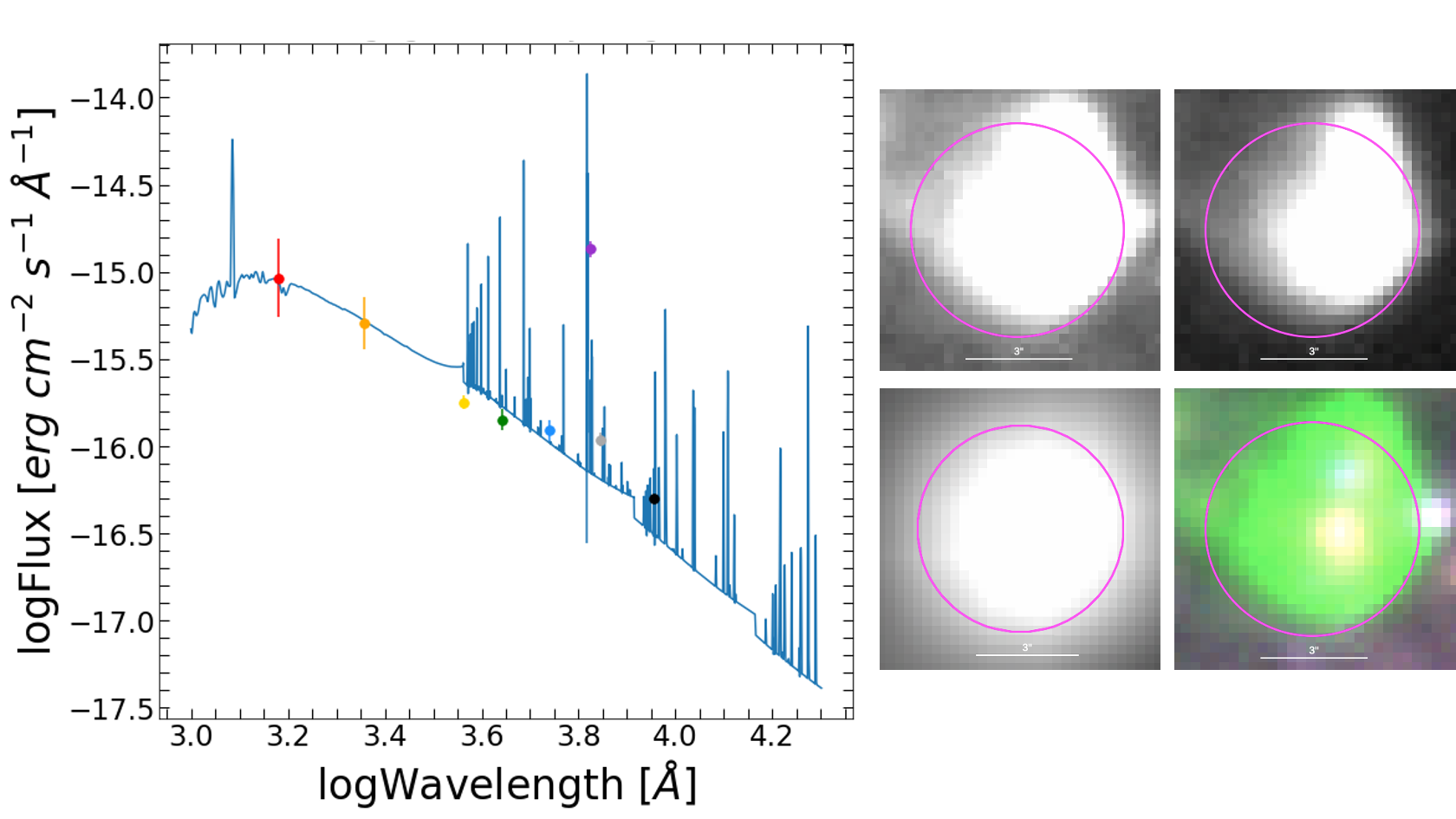}
    \caption{Cluster 58 is an examples of a $<$10~Myr cluster. Its best fits values are: 
     age = 2~Myr, mass = 890~M$_{\odot}$, and E(B-V) = 0.32~mag. The eight--band photometry with the best fits SED model (as a blue line) are shown in the left panel. The four small panels to the right show for the cluster, beginning at the top left and moving clockwise: the H$\alpha$ emission, the continuum-subtracted H$\alpha$ emission,  an  RGB color composite (red = R band, green = H$\alpha$ band, and blue = U band), and the stellar-subtracted 24$\mu$m emission. The magenta circle shows the size of the aperture used for the photometric measurements. The scale bar below the aperture represents 3". The flux scale of the images is linear, however the upper and lower limits differ between bands.}
        \label{cluster58_ex}
\end{figure}

\begin{figure}[ht]
 \centering
 \includegraphics[width=15cm]{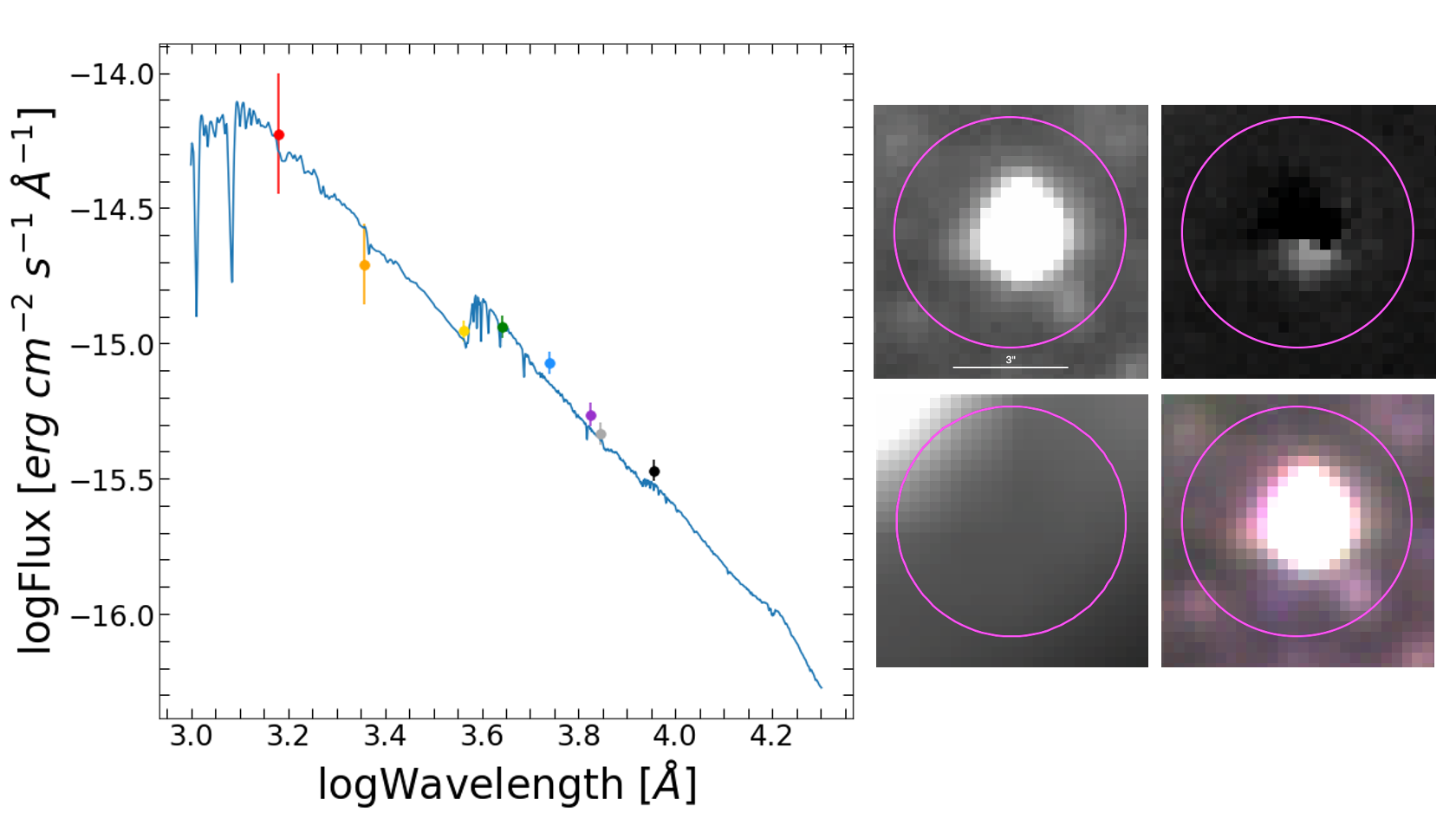}
    \caption{Cluster 118 is an example of a cluster with age between 10 and 100 Myr. Its best fits values are: 
     age = 40~Myr, mass = 11,000~M$_{\odot}$, and E(B-V) = 0.02~mag. The panels, symbols, lines, and flux scales are the same as in Fig.~\ref{cluster58_ex}.}
        \label{cluster118_ex}
\end{figure}

\begin{figure}[ht]
 \centering
 \includegraphics[width=15cm]{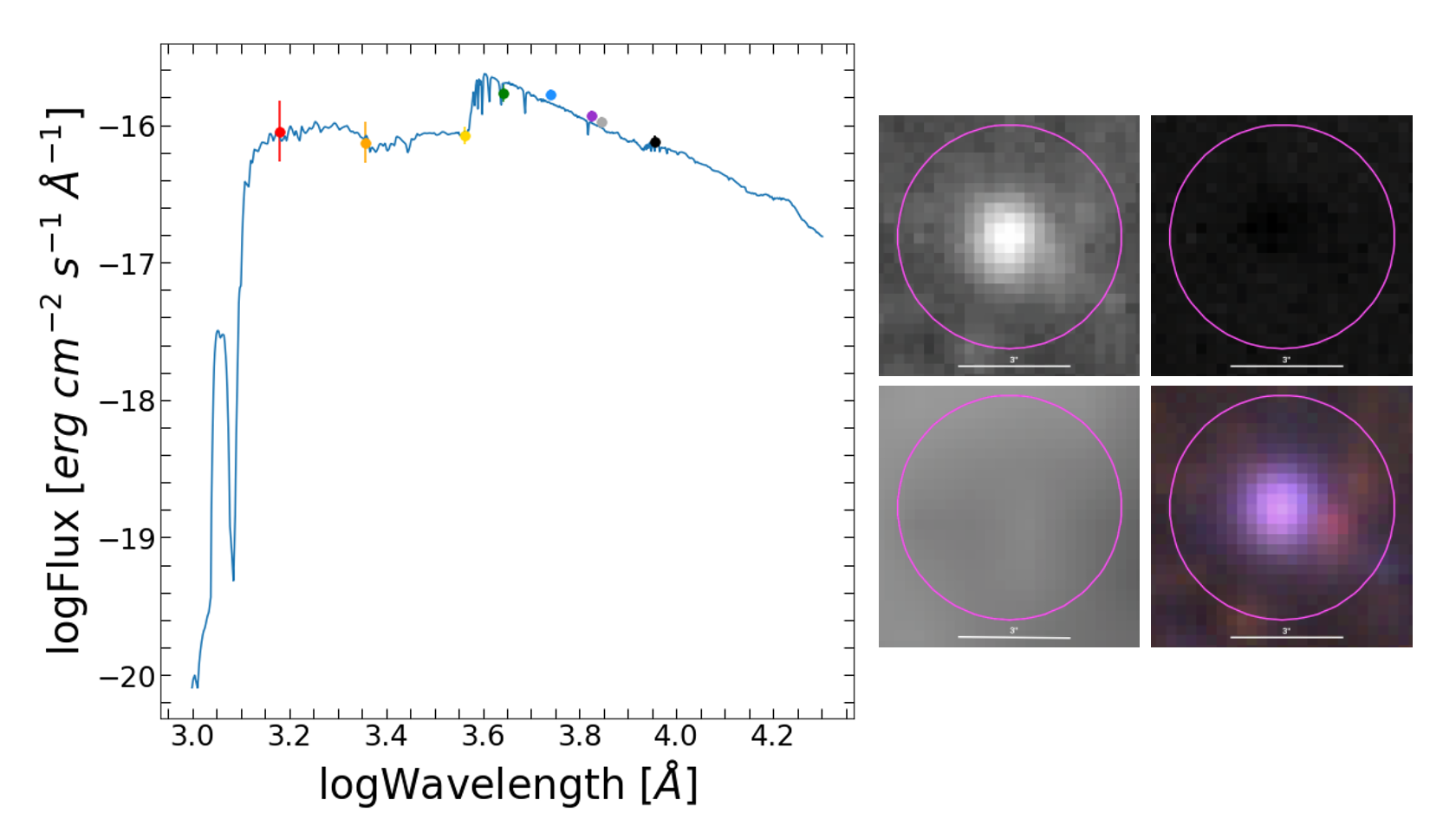}
    \caption{Cluster 124 is an example of a $>$ 100 Myr cluster. Its best fits values are: 
     age = 300~Myr, mass = 11,000~M$_{\odot}$, and E(B-V) = 0.1~mag. The panels, symbols, lines, and flux scales are the same as in Fig.~\ref{cluster58_ex}.}
        \label{cluster124_ex}
\end{figure}

\begin{figure}[ht]
 \centering
 \includegraphics[width=15cm]{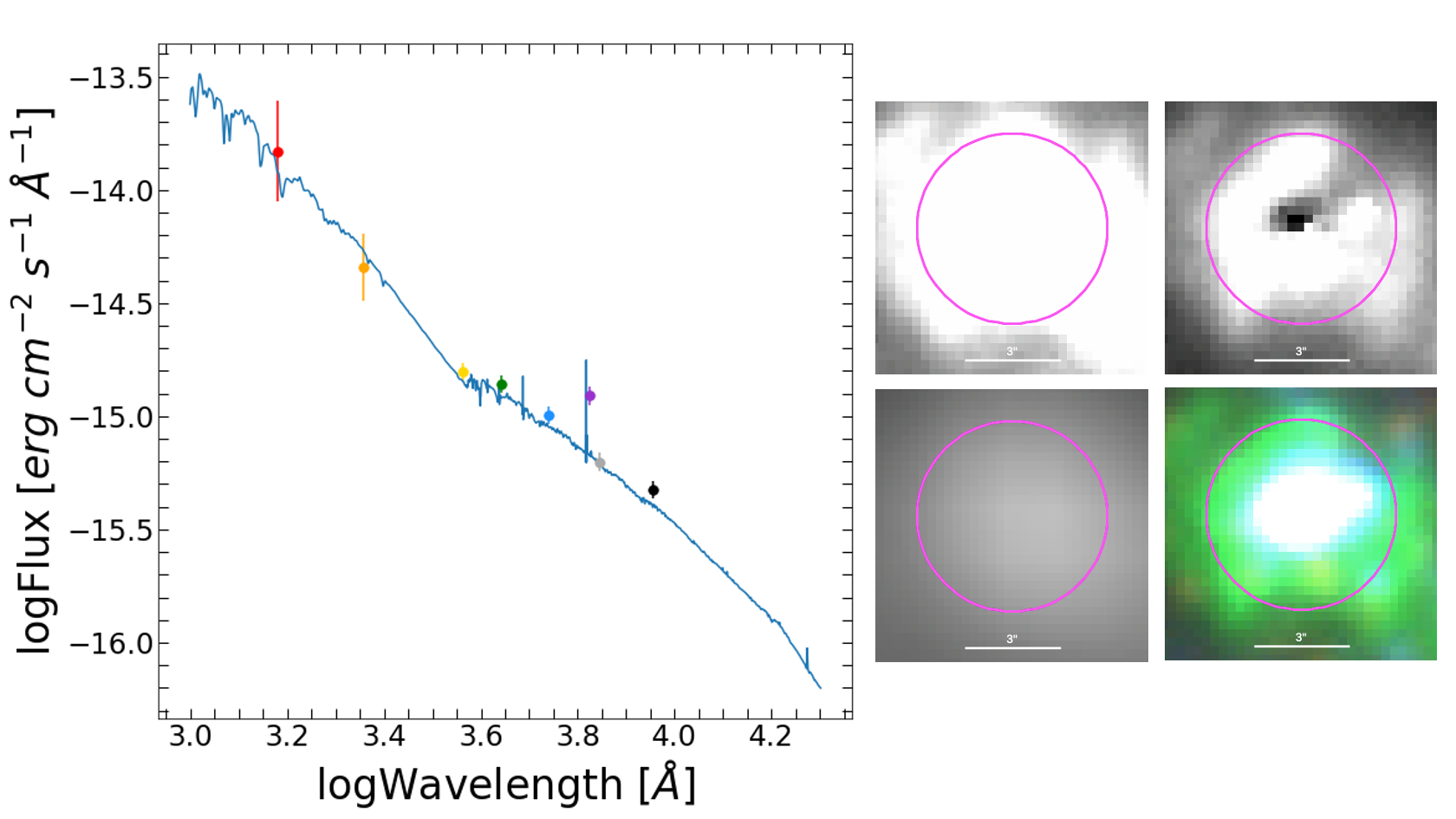}
    \caption{Cluster 103 is an example of a $<$ 10 Myr cluster with a large H$\alpha$ envelope which extends outside of the photometric aperture. Some of the H$\alpha$ emission is missed in the photometric measurements because of this, which explains some of the weakness of the H$\alpha$ emission in the SED. This cluster
    has best fit age = 8~Myr, mass = 2600~M$_{\odot}$ and E(B-V) = 0.0~mag. The panels, symbols, lines, and flux scales are the same as in Fig.~\ref{cluster58_ex}.}
        \label{cluster103_ex}
\end{figure}

The ages, masses, and extinctions of the 137 star clusters in M33 obtained from the photometry in Table~\ref{photometry} via SED fitting are listed in Table~\ref{properties}, together with the ages and masses compiled by SM2007. Examples of SEDs and their best fits are shown in Figures~\ref{LessGallery}, \ref{MidGallery}, and \ref{MoreGallery}, for five cases each with ages $\leq$10~Myr, 10~Myr--100~Myr, and $>$100~Myr, respectively. As expected, the youngest star clusters usually display H$\alpha$ in emission and strong UV fluxes relative to the optical ones, the oldest ($>$100~Myr) clusters have red SEDs with no H$\alpha$ emission, while the clusters with ages intermediate between these two often display blue SEDs, with significant  UV emission, but no or little measurable H$\alpha$ emission. A few individual cases are shown in Figures~\ref{cluster58_ex}, \ref{cluster118_ex}, and \ref{cluster124_ex} for the young, intermediate, and old stellar population cases; here we display, in addition to the observed SED and its best fit, image cutouts centered on the cluster, in the light of H$\alpha$ (both continuum--subtracted and non) and 24~$\mu$m, to show the cluster's appearance at these wavelengths. Young clusters show clear H$\alpha$ and 24~$\mu$m emission, which decreases in intensity with increasing age. In Figure~\ref{cluster103_ex}, the case of a star cluster with extended H$\alpha$ is shown; in this case, the ionized gas emission has developed a `ruptured bubble' morphology and our photometric aperture only captures a fraction of the H$\alpha$ emission from this cluster, which explains the faintness of the H$\alpha$ line in the SED plot. 

\begin{figure}[ht]
  \centering
 \includegraphics[width=8cm]{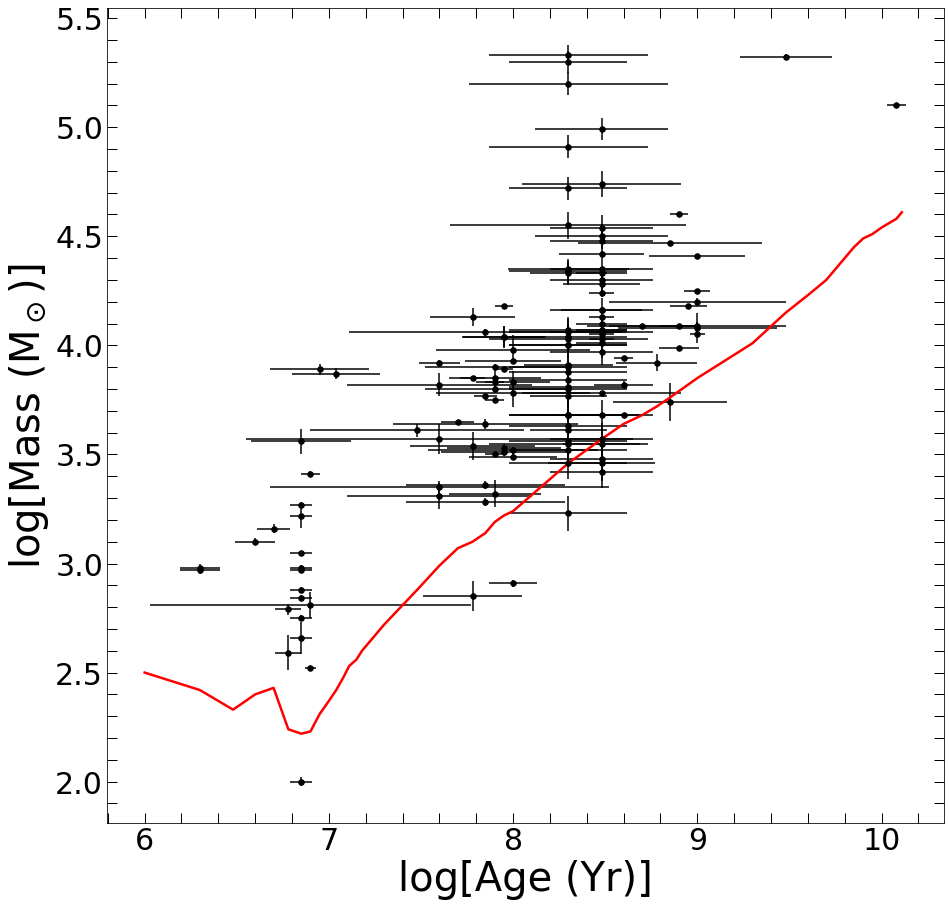}
 \includegraphics[width=9.5cm]{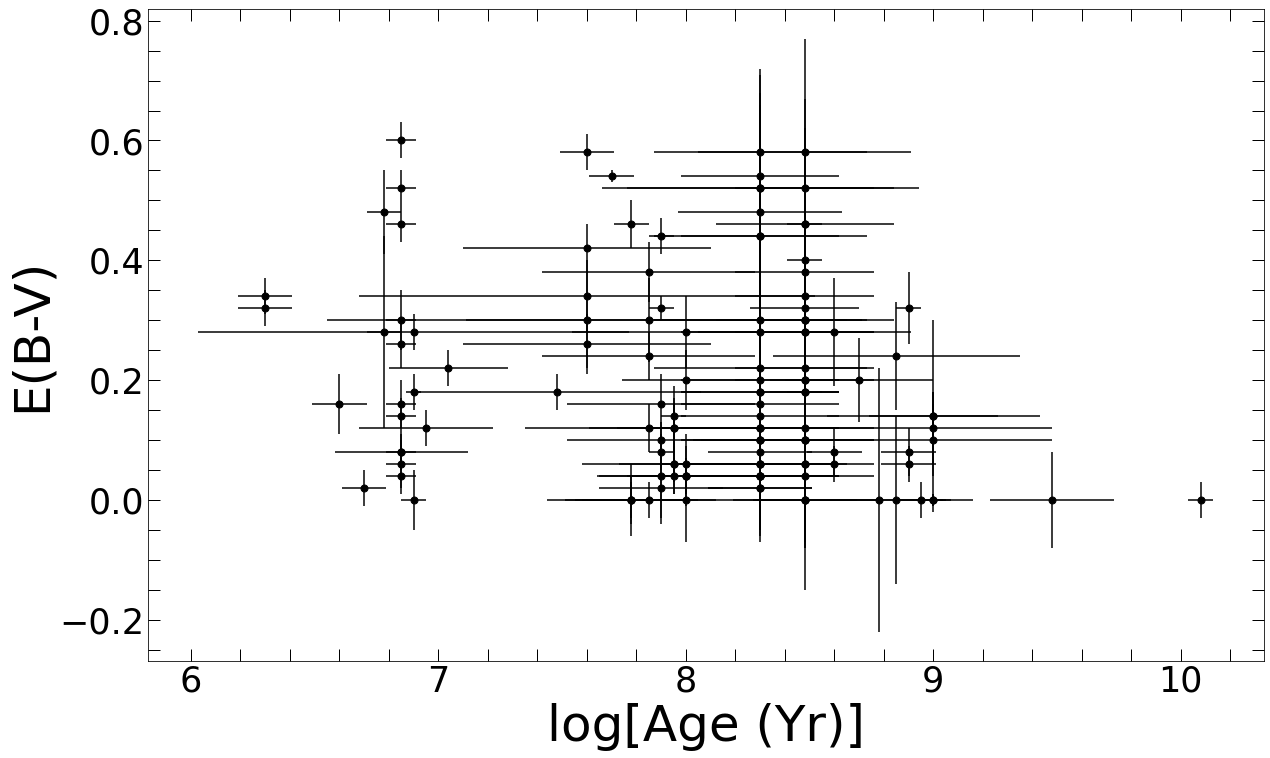}
   \caption{The masses (left panel) and color excesses E(B--V) (right panel) as a function age for the 137 star clusters in M33 derived via our SED fits. The red line in the left panel shows  the luminosity limit of the sample.}
        \label{logAge_v_logMass}
\end{figure}

The distributions of ages, masses, and extinctions are shown in graphical form in the two panels of Figure~\ref{logAge_v_logMass}, where the masses and E(B--V) are plotted as a function of the ages in the left and right panels, respectively. Statistics for the star clusters are also given in Table~\ref{cl_mass_table}.

\begin{table}[ht]
\centering   
\begin{tabular}{@{}c|c|c|c@{}}
\toprule
                            & $\leq$ 10 Myr & 10--100 Myr & $>$ 100 Myr \\ \hline
Total clusters in age bin               & 21        & 37        & 79       \\
Clusters $\ge$ 3000 M$_{\odot}$     & 2         & 30        & 75         \\
Clusters $\ge$ 5000 M$_{\odot}$     & 1         & 20        & 62         \\
\hline
\end{tabular}
    \caption{Statistics on the distribution of ages and masses for the star clusters in M33, broken into  three age bins and three mass bins. 90\% of the clusters with age $\leq$10~Myr have masses below 3,000~M$_{\odot}$, while 95\% of clusters older than 100~Myr  have masses above this value.}
    \label{cl_mass_table}
\end{table}

\begin{figure}[ht]
 \centering
 \includegraphics[width=10cm]{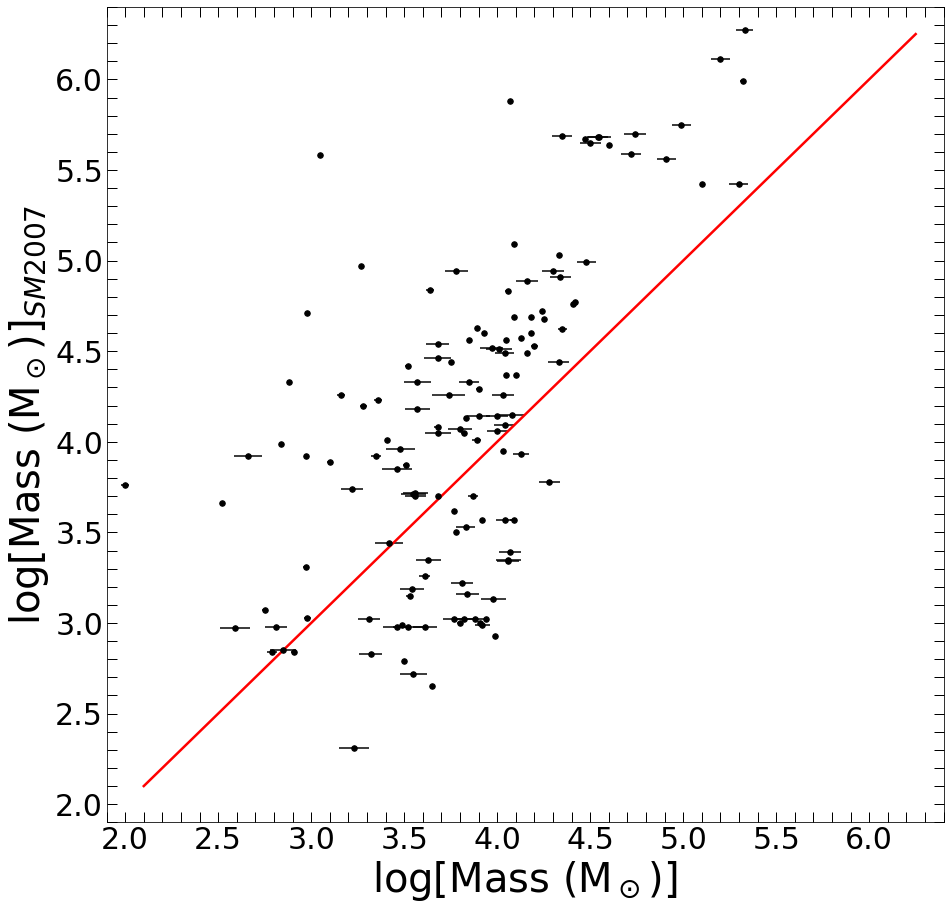}
    \caption{Comparison of our SED--fit masses with those of the SM2007 catalog. The 1--to--1 line is shown in red.}
        \label{logMass_comp}
\end{figure}

Most (90\%) of the clusters younger than 10~Myr have masses below 3,000~$M_{\odot}$ and only one is more massive than 5,000~$M_{\odot}$ (Table~\ref{cl_mass_table}), indicating that cluster masses in this age range are uncertain due to stochastic sampling of the stellar IMF \citep{Adamo+2017}. 
The general impact of stochastic  sampling on our age and mass determinations for these young clusters is that the clusters could be moderately more massive and a factor of a few older than what we find; the discrepancies between deterministic and stochastic models decrease for older ages and higher mass clusters \citep[][, their Figure~14]{Krumholz+2015}.
Older age clusters are usually  more massive, as seen in the left hand--side panel of Figure~\ref{logAge_v_logMass}, where the model line marks the standard lower luminosity limit for the detections. As expected, the masses we determine at a given age are above the model line \citep[SM2007; see, also,][]{Adamo+2017}. The trend for older clusters to be more massive than younger ones at the low mass end is simply due to fading with age, while the observation that they are more massive also at the high end of the mass distribution is the well known size--of--sample effect \citep{Hunter+2003}. Figure~\ref{logMass_comp} compares our derived masses with those reported by SM2007. The masses scatter around the 1--to--1 line (in log scale) up to $\sim$10$^4$~M$_{\odot}$, and our masses become systematically smaller than those reported in SM2007 at higher values. We attribute this discrepancy to the systematically younger ages we derive from our SED fits, which yields lower masses at a given luminosity. Overall, however, the masses derived in this work and those reported by SM2007 track each other reasonably well. The most massive clusters in this sample are about 2$\times$10$^5$~M$_{\odot}$, which is typical of spiral galaxies \citep{Larsen2009, Adamo+2018}.

\begin{figure}[ht]
 \centering
 \includegraphics[width=10cm]{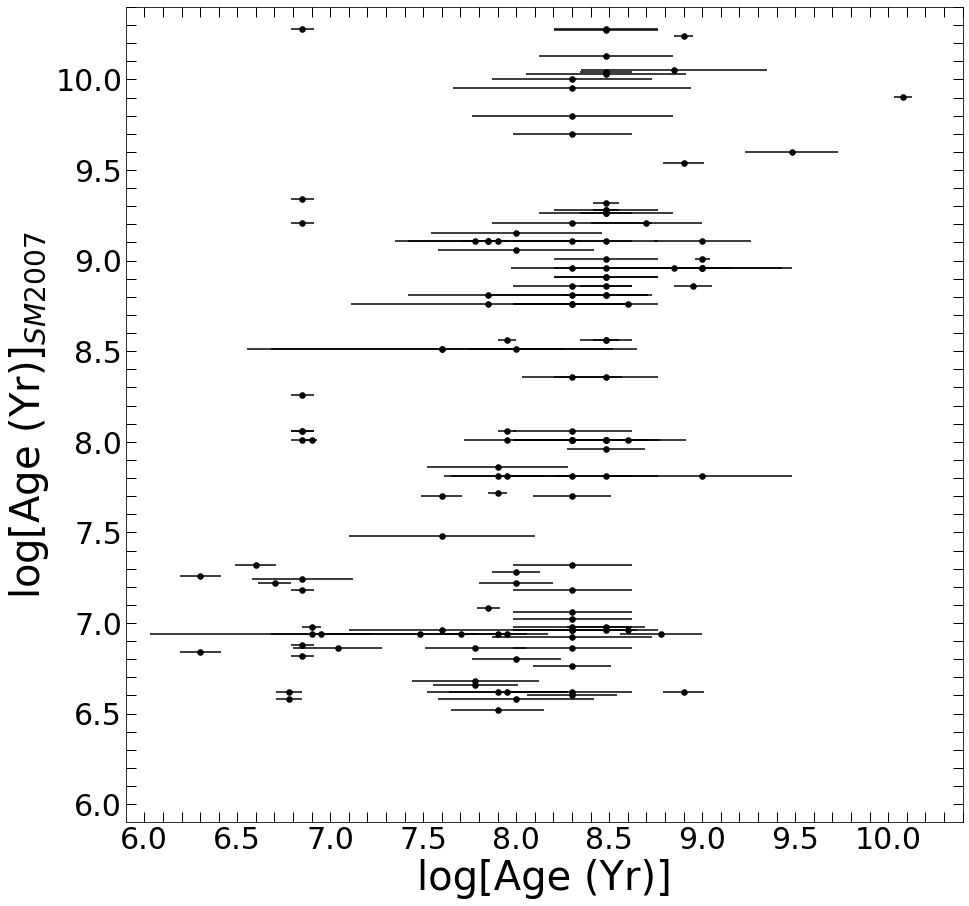}
    \caption{Comparison of our SED-fit ages with those of the SM2007 catalog.}
        \label{logAge_comp}
\end{figure}

A comparison of the ages derived here with those reported in SM2007 reveals, conversely, major discrepancies, as shown in Figure~\ref{logAge_comp}. There are several instances in which clusters that were classified as `old' (older than $\sim$100~Myr) in SM2007 are found to be around or younger than 10~Myr with our SED fits and visual checks of the H$\alpha$ images. At the opposite end, many clusters that are given ages $\lesssim$10~Myr in SM2007 are found to have much older ages by our SED fits, $>$30--40~Myr. One important characteristic of our results is that we find less than a handful of clusters with ages $>$10$^9$~yr, while SM2007 list many. \citet{Whitmore+2020} discusses in detail that there appear to be a limitation in the ability of SED fitting routines to identify old clusters correctly: in the presence of a `red' SED, fitting routines often prefer a highly--extincted, younger age solution to an extinction--free, older age solution. A work--around to this issue is to impose a limit to the highest value of E(B--V) that an older star cluster can have, in order to force an older age solution \citep[as discussed in][]{Whitmore+2020}. This approach was adopted by \citet{Ma+2004b} for a set of the M33 clusters, where they impose a constant value E(B--V)=0.1 to derive their ages. We note this limitation, which implies that several of our clusters in the 300~Myr--1~Gyr age range could be much older than what we find. 

\begin{figure}
 \centering
 \includegraphics[width=8cm]{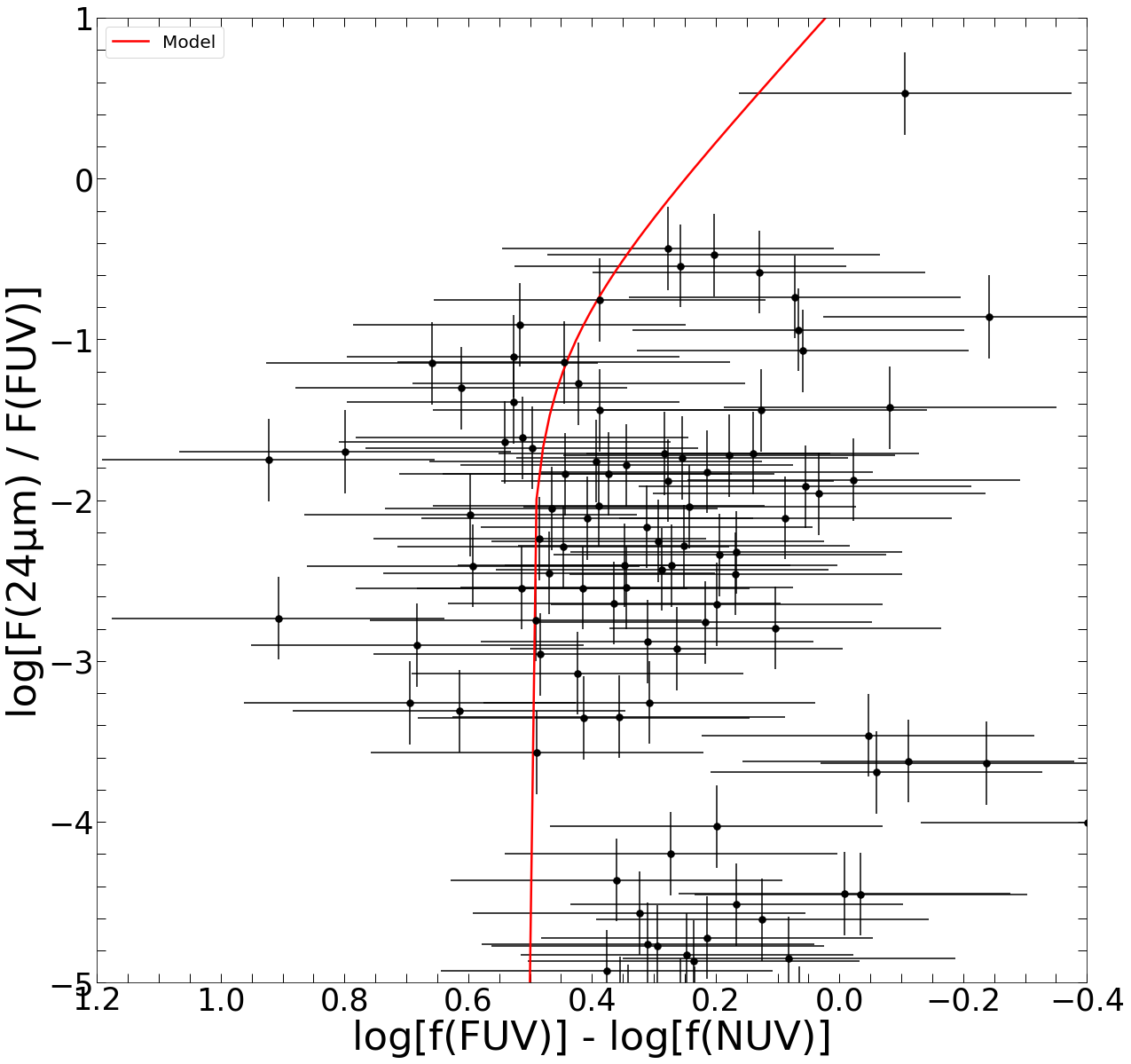}
    \caption{The flux ratio 24$\mu$m/FUV plotted against flux density ratio FUV-NUV. The y-axis is the logarithm of the ratio of the two fluxes at 24~$\mu$m and FUV, expressed as F = $\lambda f(\lambda)$ in units of $erg/cm^{2}/s$. The x-axis is the logarithm of the ratio of the flux densities in the FUV and NUV, in units of $erg/cm^{2}/$\r{A}. The 24$\mu$m/FUV ratio gives a measure of the dust extinction affecting the FUV emission \citep{Meurer+1999}. This correlation has been previously observed for galaxies and HII regions, and is extended here to clusters. The model line, in red, is from \citet{Calzetti+2005}. Most of the clusters in M33 are consistent with low--to--negligible dust attenuation. The large scatter in the FUV--NUV colors at constant 24$\mu$m/FUV value are due to the age differences in the stellar populations of the  clusters, with redder colors for older populations.}
        \label{24oFUV v FUV-NUV}
\end{figure}

Artificially `too--young' ages produce color excess values that are unusually high, or distributions that are uncharacteristically skewed towards high values. The right panel of Figure~\ref{logAge_v_logMass} shows the distribution of the best--fit E(B-V) as a function of cluster age. The maximum E(B--V) value we recover is $\lesssim$0.6~mag, and $>$2/3 of the clusters have E(B--V)$<$0.2~mag, as expected for modestly reddened environments. As a reminder, the quoted values of E(B--V) include the contribution of the foreground Milky Way dust, E(B--V)=0.036~mag. The distribution is notably `flat', with several clusters yielding ages $\ge$300~Myr and E(B--V)$\ge$0.4~mag. One such cluster is shown at the top of Figure~\ref{MoreGallery}. In this case, the UV--optical SED are not able to provide constraints on the age--dust degeneracy, and the UV emission from the stellar population is weak, implying a weak 24~$\mu$m dust emission. Figure~\ref{24oFUV v FUV-NUV} shows the distribution of the 24~$\mu$m/FUV flux ratio as a function of the FUV--NUV color. The  24~$\mu$m/FUV flux ratio measures the fraction of UV light absorbed by dust and re--emitted in the infrared; the FUV--NUV color, here given as the ratio of the flux density at the two wavelengths, measures the amount of dust reddening suffered by the UV SED \citep{Meurer+1999, Calzetti+2000}. The star clusters in M33 are consistent with modest--to--negligible  reddening and processing of UV light into the infrared, in general agreement with our findings for the E(B--V) values. The red (more negative) FUV--NUV colors at very low values of the 24~$\mu$m/FUV ratio are easily explained with effects of age in the clusters SEDs \citep{Calzetti+2005, Cortese+2006}. From this analysis, we conclude that, while many of the  clusters with ages $\ge$300~Myr in our sample could be older if they are less dusty than what we find from our SED fits \citep[see discussion in][]{Whitmore+2020}, the data at hand do not allow us to unequivocally conclude this. However, our age determinations for the clusters younger than $\approx$300~Myr are reasonably robust. 

\subsection{Colors versus Age and Extinction}

\begin{figure}
 \centering
 \includegraphics[width=14cm]{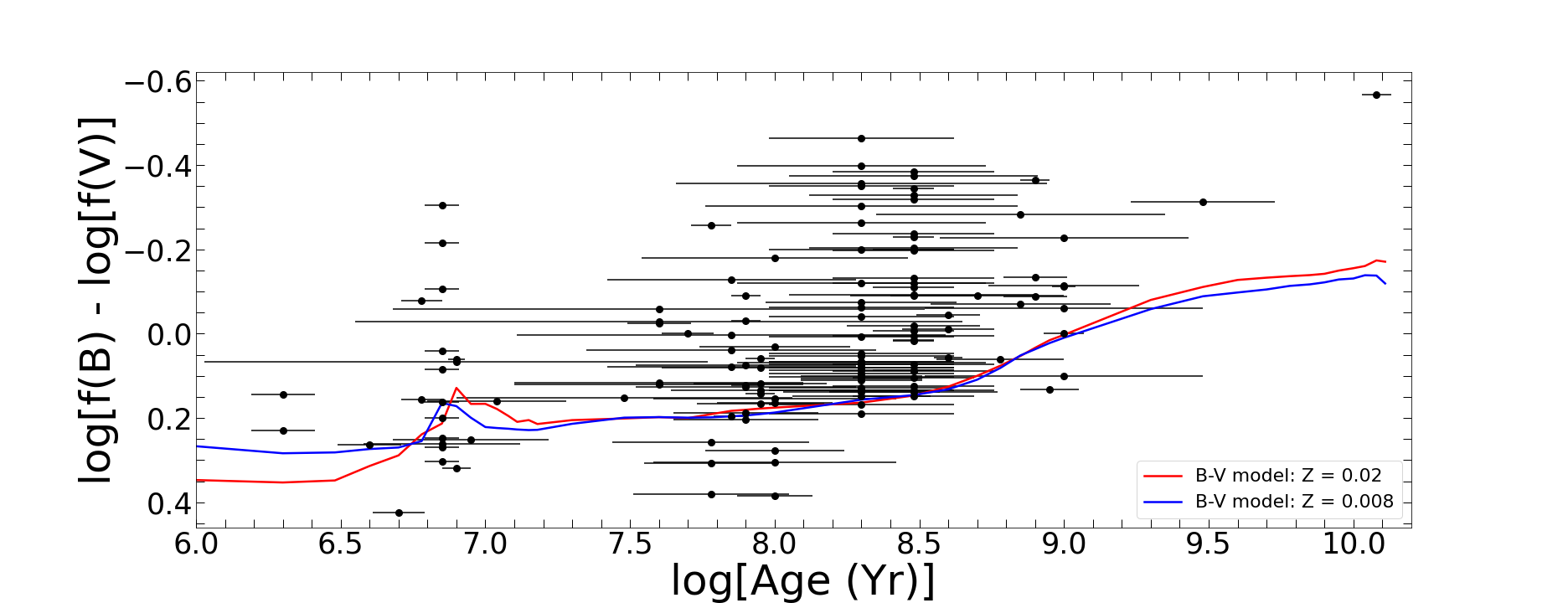}
    \caption{The B--V flux ratio v. age  using our measurements, without any extinction correction. Correcting the flux ratio for the effect of  the Milky Way foreground extinction would make it bluer (move downward) by 0.036~dex, which is significantly smaller than the scatter in the data.}
        \label{B-VAge}
\end{figure}

\begin{figure}
 \centering
 \includegraphics[width=10cm]{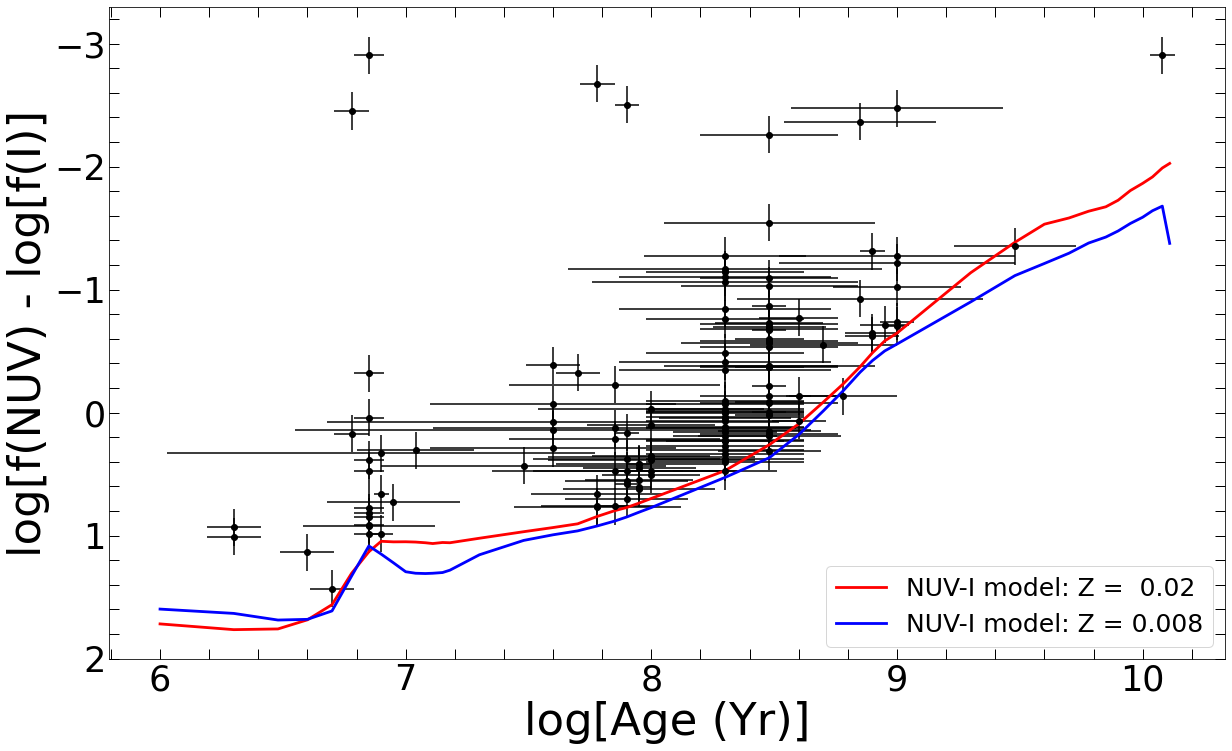}
    \caption{NUV-I prior to extinction corrections plotted against our calculated ages. For the FUV--I color axis, blue is down and red is up. Correction for the Milky Way foreground extinction makes the color bluer by less than 0.1~dex.}
        \label{NUV-I}
\end{figure}

\begin{figure}
 \centering
 \includegraphics[width=10cm]{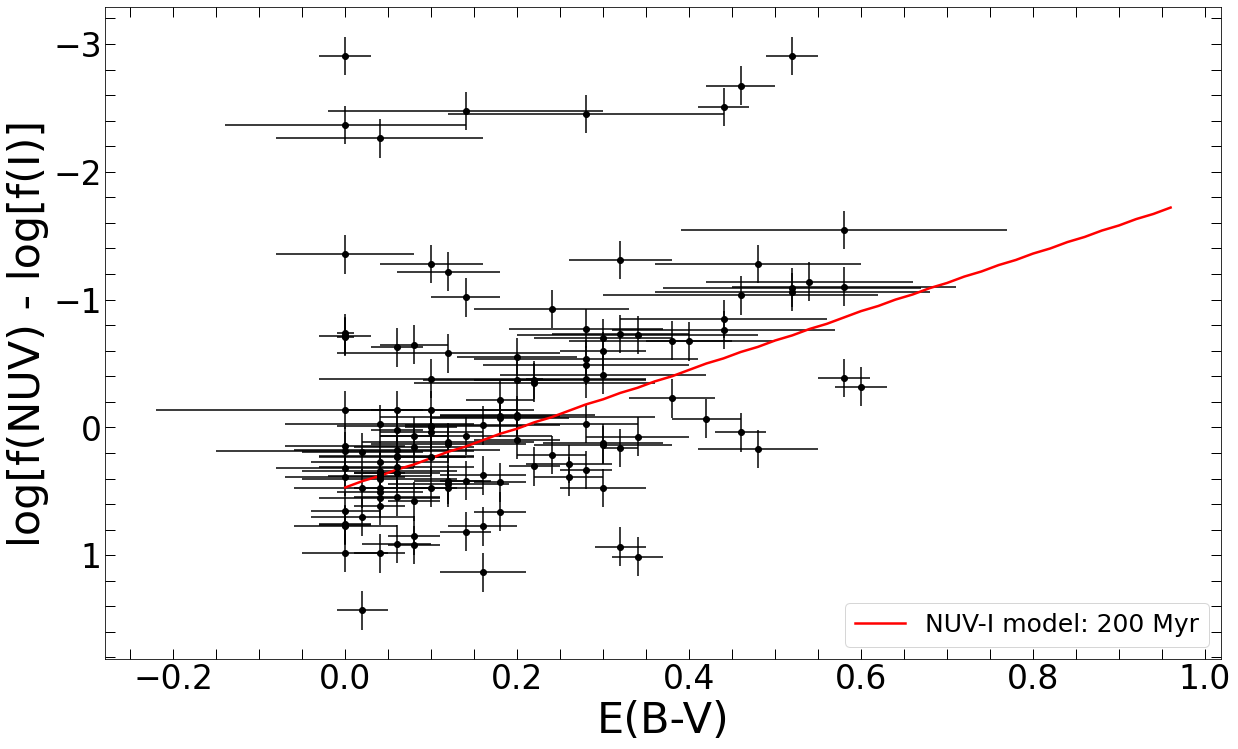}
    \caption{NUV-I without extinction corrections plotted against the measured E(B-V) values. The over-plotted model is set at the median age of the dataset, 200 Myr.}
        \label{NUV-I v E(B-V)}
\end{figure}

The B--V color becomes redder for increasing age, as discussed by many authors, including SM2007. Our measurements follow this general trend, although we note a significant scatter towards red colors at fixed age (Figure~\ref{B-VAge}). A large scatter in the direction of redder colors at fixed age is also observed for the NUV--I color, although there is still a general trend of redder colors for increasing age  (Figure~\ref{NUV-I}). This color, and similar colors (e.g., NUV--K) that use a large wavelength baseline to magnify effects, was introduced as an age indicator for galaxies once the effects of internal extinction are removed \citep[e.g.,][]{Munoz+2007, Cortese+2008}. We verify here that the same colors can be used as effective age indicators also in the case of star clusters. As already discussed by other authors, one downside of the long wavelength baseline is the sensitivity of the NUV--I color to dust extinction, as shown both by the scatter noted above (Figure~\ref{NUV-I}) and by the trend of redder colors for increasing E(B--V) values we observe for our clusters (Figure~\ref{NUV-I v E(B-V)}). In this figure we show the locus of a 200~Myr  star cluster for increasing values of E(B--V) as a red line. Most clusters  in our sample scatter around this line, with $<$10 clusters deviating from this trend by showing much redder NUV--I colors. 

In Figures~\ref{B-VAge} and \ref{NUV-I} we show model expectations for both metallicities Z = 0.02 (red lines) and Z = 0.008 (blue lines). Notably, the differences between the two model lines are small, and well within our measurement and parameter fit uncertainties;  this further justifies the use of Z = 0.02 models for our SED fits. As already mentioned above, while the colors follow the general trend of the age lines, the data scatter mostly {\em above} these lines; this is particularly evident for the NUV--I color (Figure~\ref{NUV-I}), and is the expected effect of dust extinction.  

\section{Summary and Conclusions} \label{sec:conclusions}

Leveraging the extended wavelength baseline provided by the \textit{GALEX} UV together with optical and H$\alpha$ bands, we have used SED--fitting to derive improved ages, masses, and extinction values for 137 confirmed star clusters in M33. Identified from the compilation of \citet{Sarajedini2007}, all clusters also have masses and ages derived in a series of papers by \citet{Ma+2001,  Ma+2002b, Ma+2002c, Ma+2002a, Ma+2004a, Ma+2004b}. The 24~$\mu$m emission detected by the \textit{Spitzer Space Telescope} provides an additional discriminant for breaking the age--extinction degeneracy below 100~Myr. We exclude 26 clusters from the original list of 163 for a number of reasons that affect their SEDs, such as presence of multiple sources with different colors within the measurement aperture (Table~\ref{exclusions}). We compare our results with the earlier ones. 

In terms of cluster ages, we find overall agreement with the earlier results, but also noticeable differences. About half dozen clusters that were given ages older than 100~Myr are younger than $\sim$10~Myr based on our method. We also find ages $\gtrsim$100~Myr for about a dozen clusters that were previously given ages$\lesssim$10~Myr. We note a dearth of clusters with ages $>$1~Gyr in our sample, possibly due to a well--known problem with  SED fits which tend to prefer younger ages at higher extinctions over older, less extincted solutions \citep{Whitmore+2020}. As a result, we observe a clumping of  clusters around $\sim$200--300~Myr of age.

Unlike ages, our SED--derived masses track the earlier determinations reasonably well. The main discrepancy is that we do not find cluster masses $\gtrsim$2$\times$10$^5$~M$_{\odot}$, while SM2007 reports a little over a dozen clusters above this mass. We attribute much of the discrepancy to the dearth of clusters older than $\sim$1~Gyr in our fits. We should remark that the mass range we find for the star clusters in M33 is consistent with the mass range found for star clusters in other spiral galaxies at comparable age intervals \citep{Adamo+2018}. 

All our star clusters have moderate--to--small internal extinction, all with E(B--V)$<$0.6~mag, and about 2/3 with E(B--V)$<$0.2~mag. We do not observe a trend between age and extinction, although the fast timescale for the diffusion of star clusters in galaxies \citep{grasha2017a, grasha2017b} would imply that they reach regions of low dust content within several  tens of Myr. Aforementioned age--extinction degeneracies for intermediate age clusters ($\ge$200--300~Myr) could explain why we observe an extinction value as high as $\sim$0.6~mag in a $\sim$300~Myr old cluster; an older age with a lower extinction may be an alternate, viable explanation.  

In addition to providing the list of physical parameters for the 137 clusters whose SEDs  we fit, we also publish our  10--band photometry for all 163 clusters, which includes: \textit{GALEX} FUV and NUV, optical U, B, V, R, H$\alpha$, and I, and infrared centered at 3.6~$\mu$m and 24~$\mu$m. Our photometry table can provide a resource to test alternative methods for deriving clusters' physical parameters that may overcome some of the limitations discussed in this work.

\appendix

\begin{longrotatetable}
\begin{deluxetable*}{p{1cm}p{1cm}p{1.5cm}p{1.5cm}p{0.7cm}p{0.5cm}p{1cm}p{1cm}p{0.5cm}p{1cm}p{0.5cm}p{1cm}p{5cm}}
\tablecaption{Derived and literature physical properties of the M33 star clusters\label{properties}}
\tablewidth{500pt}
\tabletypesize{\scriptsize}
\tablehead{
\colhead{SM2007} & \colhead{ID} & 
\colhead{RA} & \colhead{Dec} & 
\colhead{SM2007} & \colhead{Age} & 
\colhead{Age Err} & \colhead{SM2007} & 
\colhead{Mass} & \colhead{Mass Err} & \colhead{E(B-V)} & \colhead{E(B-V) Err} & \colhead{Comments} \\
                \colhead{ID} & \colhead{This Work} & \colhead{} & \colhead{} & \colhead{Age} & \colhead{} & \colhead{} & \colhead{Mass} & \colhead{} & \colhead{} & \colhead{} & \colhead{} & \colhead{}
}
\startdata
\centering{9} & \centering{0} & \centering{ 01 32 42.93} & \centering{+30 35 38.6} & \centering{8.56} & \centering{6.85} & \centering{$\pm$0.06} & \centering{4.63} & \centering{3.06} & \centering{$\pm$0.082} & \centering{0.3} & \centering{ $\pm$0.08} & \\
\centering{10} & \centering{1} & \centering{ 01 32 44.30} & \centering{+30 40 12.4} & \centering{10.28} & \centering{6.85} & \centering{$\pm$0.06} & \centering{5.58} & \centering{3.05} & \centering{$\pm$0.025} & \centering{0.6} & \centering{ $\pm$0.07}& \\
\centering{16} & \centering{2} & \centering{ 01 32 52.65} & \centering{+30 14 30.9} & \centering{8.01} & \centering{8.3} & \centering{$\pm$0.32} & \centering{3.72} & \centering{3.56} & \centering{$\pm$0.071} & \centering{0.04} & \centering{ $\pm$0.08}& \\
\centering{26} & \centering{3} & \centering{ 01 32 56.32} & \centering{+30 14 58.9} & \centering{10.04} & \centering{8.48} & \centering{$\pm$0.07} & \centering{5.88} & \centering{4.05} & \centering{$\pm$0.006} & \centering{0.04} & \centering{ $\pm$0.03}& \\
\centering{40} & \centering{4} & \centering{ 01 33 08.11} & \centering{+30 28 00.2} & \centering{9.15} & \centering{8.3} & \centering{$\pm$0.44} & \centering{4.42} & \centering{3.63} & \centering{$\pm$0.024} & \centering{0.18} & \centering{ $\pm$0.08}& \\
\centering{41} & \centering{5} & \centering{ 01 33 09.82} & \centering{+30 12 50.7} & \centering{7.96} & \centering{8.3} & \centering{$\pm$0.43} & \centering{3.71} & \centering{3.5} & \centering{$\pm$0.08} & \centering{0.06} & \centering{ $\pm$0.07}& \\
\centering{42} & \centering{6} & \centering{ 01 33 10.11} & \centering{+30 29 56.9} & \centering{6.58} & \centering{6.78} & \centering{$\pm$0.07} & \centering{2.97} & \centering{2.57} & \centering{$\pm$0.064} & \centering{0.26} & \centering{ $\pm$0.12}& \\
\enddata
\tablecomments{SED--fitting derived ages, masses, and colors excesses,  E(B--V), together with 1~$\sigma$ uncertainties for 137 confirmed star clusters in M33, listed together with the ages and masses compiled in \citet[][, indicated as SM2007]{Sarajedini2007}. The E(B--V) values in this Table are the sum of the internal dust reddening in the star cluster plus the foreground reddening of the Milky Way, E(B--V)=0.036 \citep{Schlafly+2011}. In some instances of two or more sources present in the aperture, the best fit results are retained if the sources appear of comparable age (similar behavior in colors), as indicated in the Comments field; in this case, the quoted mass is the sum of all masses for the sources in the aperture. For sources yielding with two best--fits to the age/extinction, the reported value reflects an assessment based on the 24~$\mu$m emission. The retained best fit is given in the Comments field. }
\end{deluxetable*}
\end{longrotatetable}

\begin{longrotatetable}
\begin{deluxetable*}{p{1cm}p{1cm}p{1.5cm}p{1.5cm}p{1.5cm}p{1.5cm}p{1.5cm}p{1.5cm}p{1.5cm}p{1.5cm}p{1.5cm}p{1.5cm}p{0cm}}
\tablecaption{Photometric measurements of the M33 star clusters\label{photometry}}
\tablewidth{500pt}
\tabletypesize{\scriptsize}
\tablehead{\colhead{ID} & \colhead{ID} & \colhead{f(FUV)} & \colhead{f(NUV)} & \colhead{f(U)} & \colhead{f(B)} & \colhead{f(V)} & \colhead{f(R)} & \colhead{f(H$\alpha$)} & \colhead{f(I)} & \colhead{F(3.6$\mu$m)} & \colhead{F(24$\mu$m)} & \\
               \colhead {SM2007} & \colhead {This Work} & \colhead{\tiny$erg/s/cm^{2}/\r{A}$} & \colhead{\tiny$erg/s/cm^{2}/\r{A}$} & \colhead{\tiny$erg/s/cm^{2}/\r{A}$} & \colhead{\tiny$erg/s/cm^{2}/\r{A}$} & \colhead{\tiny$erg/s/cm^{2}/\r{A}$} & \colhead{\tiny$erg/s/cm^{2}/\r{A}$} & \colhead{\tiny$erg/s/cm^{2}/\r{A}$} &
               \colhead{\tiny$erg/s/cm^{2}/\r{A}$} &
                \colhead{\tiny$erg/s/cm^{2}$} & \colhead{\tiny$erg/s/cm^{2}$} &
} 
\startdata
\centering{9} & \centering{0} & \centering{4.43E-16} & \centering{2.79E-16} & \centering{1.52E-16} & \centering{2.48E-16} & \centering{2.34E-16} & \centering{1.54E-16} & \centering{2.01E-16} & \centering{1.07E-16} & \centering{1.56E-13} & \centering{$<$4.3E-16} & \\ 
\centering{10} & \centering{1} & \centering{2.88E-17} & \centering{2.54E-17} & \centering{2.4E-17} & \centering{5.09E-17} & \centering{6.9E-17} & \centering{5.84E-17} & \centering{6.91E-17} & \centering{5.29E-17} & \centering{8.79E-14} & \centering{2.67E-15} & \\
\centering{16} & \centering{2} & \centering{1.4E-16} & \centering{7.83E-17} & \centering{6.34E-17} & \centering{1.01E-16} & \centering{9.39E-17} & \centering{5.81E-17} & \centering{6.4E-17} & \centering{3.6E-17} & \centering{2.67E-14} & \centering{5.55E-15} & \\
\centering{26} & \centering{3} & \centering{1.63E-16} & \centering{9.12E-17} & \centering{1.37E-16} & \centering{2.4E-16} & \centering{2.2E-16} & \centering{1.35E-16} & \centering{1.49E-16} & \centering{8.72E-17} & \centering{5.63E-14} & \centering{$<$4.3E-16} & \\
\centering{40} & \centering{4} & \centering{5.17E-17} & \centering{2.89E-17} & \centering{2.62E-17} & \centering{4.9E-17} & \centering{5.87E-17} & \centering{4.32E-17} & \centering{4.51E-17} & \centering{3.09E-17} & \centering{3.36E-14} & \centering{7.23E-15} & \\
\centering{41} & \centering{5} & \centering{1.06E-16} & \centering{4.19E-17} & \centering{6.11E-17} & \centering{1.04E-16} & \centering{8.96E-17} & \centering{5.48E-17} & \centering{5.25E-17} & \centering{2.99E-17} & \centering{1.4E-14} & \centering{6.27E-15} & \\
\centering{42} & \centering{6} & \centering{$<$7.8E-19} & \centering{$<$9.4E-19} & \centering{7.48E-17} & \centering{1.06E-16} & \centering{9.07E-17} & \centering{5.73E-17} & \centering{1.26E-16} & \centering{3.66E-17} & \centering{3.47E-15} & \centering{6.63E-14} & \\
\enddata
\tablecomments{Photometry in 10 bands for 163 confirmed M33 star  clusters in the compilation of \citet{Sarajedini2007}. The values for the U, B, V, R, H$\alpha$, and I bands are flux densities in units of erg~s$^{-1}$~cm$^{-2}$~\r{A}$^{-1}$, while the values for the IR bands, 3.6~$\mu$m and 24~$\mu$m, are fluxes in units of erg~s$^{-1}$~cm$^{-2}$. The flux densities for the UV and optical bands are reported as directly measured, i.e., {\em no correction} for the foreground Milky Way extinction has been applied. Uncertainties on the measurements are listed in Table~\ref{logf_err_table}}.
\end{deluxetable*}
\end{longrotatetable}

\begin{longrotatetable}
\begin{deluxetable*}{p{1cm}p{1cm}p{1.5cm}p{1.5cm}p{2cm}p{2cm}p{8.5cm}}
\tablecaption{Clusters excluded from analysis.\label{exclusions}}
\tablewidth{500pt}
\tabletypesize{\scriptsize}
\tablehead{
\colhead{SM2007 ID} & \colhead{Cluster ID} & 
\colhead{RA} & \colhead{Dec} & 
\colhead{SM2007 Age} & \colhead{SM2007 Mass} & 
\colhead{Reason for Exclusion} \\ 
} 
\startdata
\centering{44} & \centering{7} & \centering{ 01 33 13.80} & \centering{+30 29 03.6} & \centering{10.06} & \centering{5.32} & {Three sources within aperture, inspection of these sources shows they have different ages (brightest at different wavelengths).} \\
\centering{66} & \centering{13} & \centering{ 01 33 22.11} & \centering{+30 40 28.4} & \centering{9.11} & \centering{4.73} & {Two--three sources within aperture, inspection of these sources shows they have different ages.} \\
\centering{67} & \centering{14} & \centering{ 01 33 22.16} & \centering{+30 40 26.0} & \centering{6.94} & \centering{3.04} & {Secondary source within aperture of different age.} \\ 
\centering{83} & \centering{21} & \centering{ 01 33 26.49} & \centering{+30 41 11.6} & \centering{6.6} & \centering{2.62} & {Source is dim, SED data is unusual and noisy.} \\ 
\centering{90} & \centering{26} & \centering{ 01 33 28.00} & \centering{+30 21 06.2} & \centering{6.84} & \centering{2.91} & {Multiple sources within aperture of different ages.} \\ 
\centering{94} & \centering{29} & \centering{ 01 33 28.70} & \centering{+30 36 37.5} & \centering{8.01} & \centering{4.35} & {Multiple sources within aperture of different ages.} \\
\centering{113} & \centering{37} & \centering{ 01 33 32.01} & \centering{+30 33 21.8} & \centering{6.98} & \centering{3.54} & {Multiple sources within aperture of different ages.} \\ 
\centering{116} & \centering{39} & \centering{ 01 33 32.36} & \centering{+30 38 28.0} & \centering{9.95} & \centering{5.25} & {Source is dim, secondary source in the aperture, and bleeding from nearby sources.} \\
\enddata
\end{deluxetable*}
\end{longrotatetable}

\end{document}